# Detecting and quantifying palaeoseasonality in stalagmites using geochemical and modelling approaches

James U.L. Baldini[1], Franziska A. Lechleitner[2,3], Sebastian F.M. Breitenbach[4], Jeroen van Hunen[1], Lisa M. Baldini[5], Peter M. Wynn[6], Robert A. Jamieson[7], Harriet E. Ridley[1], Alex J. Baker[8], Izabela W. Walczak[9], and Jens Fohlmeister[10,11]

[1]Department of Earth Sciences, Durham University, DH1 3LE, United Kingdom.

[2]Department of Earth Sciences, University of Oxford, South Parks Road, Oxford OX1 3AN, United Kingdom.

[3]Laboratory for the Analysis of Radiocarbon with AMS (LARA), Department of Chemistry and Biochemistry, and Oeschger Centre for Climate Change Research, University of Bern, Freiestrasse 3, 3012 Bern, Switzerland.

[4]Department of Geography and Environmental Sciences, Northumbria University, Newcastle upon Tyne, NE1 8ST, United Kingdom.

[5]School of Health & Life Sciences, Teesside University, Middlesbrough, TS1 3BX, United Kingdom

[6]Lancaster Environment Centre, Lancaster University, Lancaster, LA1 4YQ, United Kingdom.

[7]School of Earth and Environment, University of Leeds, Leeds, LS2 9JT, United Kingdom

[8]National Centre for Atmospheric Science and Department of Meteorology, University of Reading, RG6 6BB, United Kingdom.

[9]Durham Centre for Academic Development, Durham University, Durham, DH1 1TA, United Kingdom.

[10]Potsdam Institute for Climate Impact Research, Telegrafenberg, 14473 Potsdam, Germany.

[11]GFZ German Research Centre for Geosciences, Section 'Climate Dynamics and Landscape Development', Telegrafenberg, 14473 Potsdam, Germany.

**Abstract**

Stalagmites are an extraordinarily powerful resource for the reconstruction of climatological palaeoseasonality. Here, we provide a comprehensive review of different types of seasonality preserved by stalagmites and methods for extracting this information. A new drip classification scheme is introduced, which facilitates the identification of stalagmites fed by seasonally responsive drips and which highlights the wide variability in drip types feeding stalagmites. This hydrological variability, combined with seasonality in Earth atmospheric processes, meteoric precipitation, biological processes within the soil, and cave atmosphere composition means that every stalagmite retains a different and distinct (but correct) record of environmental conditions. Replication of a record is extremely useful but should not be expected unless comparing stalagmites affected by the same processes in the same proportion. A short overview of common microanalytical techniques is presented, and suggested best practice discussed. In addition to geochemical methods, a new modelling technique for extracting meteoric precipitation and temperature palaeoseasonality from stalagmite $\delta^{18}O$ data is discussed and tested with both synthetic and real-world datasets. Finally, world maps of temperature, meteoric precipitation amount, and meteoric precipitation oxygen isotope ratio seasonality are presented and discussed, with an aim of helping to identify regions most sensitive to shifts in seasonality.

**1. Introduction**

Over the past few decades stalagmites have become one of the most important terrestrial archives of climate and environmental change. Their widespread distribution, amenability to radiometric dating, and capacity for retaining seasonal- to decadal-scale environmental information have made them indispensable archives for a wide variety of climate information, most commonly rainfall or temperature variability. The field has developed rapidly, and it is now clear that stalagmites generally do not record a single climate parameter (e.g., cave temperature, rainfall amount, etc.) exclusively, but instead record a combination of processes.  It is increasingly acknowledged that every stalagmite contains a robust history of some aspect of environmental change. The issue is one of complexity; generally speaking, the stalagmite with the least complex signal is considered the ideal. Records generated from stalagmites with more complex stratigraphies, whose drip flow route changes through time, or that are influenced by numerous environmental processes, often prove more difficult to interpret. Some stalagmite records may miss short-lived climate excursions because they are fed by drips that do not respond to the transient climate forcing in question. Others might lose sensitivity or respond non-linearly to a climate forcing; for example, a stalagmite might record droughts faithfully, but miss exceptionally wet intervals when the epikarst (the highly fractured transition zone between soil and bedrock) is saturated with water. To exacerbate the issue further, most published stalagmite records lack the requisite analytical resolution to detect palaeoseasonality, an aspect of the climate signal that is increasingly recognised as critical to the interpretation of geochemical records from stalagmites (Baldini et al., 2019; Morellón et al., 2009; Moreno et al., 2017). In other words, the desired climate signal is often compromised by: i) inherent complexities associated with the hydrological transfer of the climate signal to the stalagmite, ii) overprinting of the desired climate-driven signal by other environmental variables, and iii) bias introduced via the necessarily selective sampling of the stalagmite for analysis. The challenge for palaeoclimatologists is to extract and correctly interpret the desired climate signal from a stalagmite, bearing these complexities in mind.

The detection of a seasonality signal within a stalagmite can greatly help interpret all datasets from a stalagmite sample, of any temporal resolution. For example, the detection of a seasonal geochemical cycle can contribute to chronological models (Baldini et al., 2002; Carlson et al., 2018; Ridley et al., 2015b), in some cases permitting the development of high-precision chronologies over extended time intervals (Ban et al., 2018; Carlson et al., 2018; Duan et al., 2015; Nagra et al., 2017; Ridley et al., 2015b; Smith et al., 2009). Unlike most other laminated records (e.g., tree rings, ice cores), high-precision radiometric dates can anchor stalagmite layer count chronologies, reducing accumulated counting errors. Proxy information from laminated stalagmites can be linked to environmental variability at seasonal resolution (Mattey et al., 2010; Orland et al., 2019; Ridley et al., 2015b), allowing much needed insights into past climatic dynamics that are difficult to obtain otherwise.

The fact that stalagmites can reveal palaeoseasonality, a notoriously difficult climate parameter to reconstruct, is critical for identifying wholesale shifts in climate belts. For example, monthly-scale geochemical data from a stalagmite has detected variability in the Intertropical Convergence Zone influence on rainfall seasonality in Central America over the last two millennia (Asmerom et al., 2020) and the shift from a maritime to a more continental climate in western Ireland in the early Holocene (Baldini et al., 2002), transitions which must otherwise be inferred using annual- to centennial-resolution data (e.g., Breitenbach et al., 2019). High spatial resolution approaches yielding palaeoseasonality can distinguish rainfall occurring at different times of the year, for example, monsoonal rainfall versus dry season rainfall (Ban et al., 2018; Ronay et al., 2019), providing a wealth of information unattainable by other means.

Seasonality is one of climate's most important aspects, and this is reflected in the basic subdivisions of the Köppen system, the most commonly used climate classification scheme (Köppen, 1918; Peel et al., 2007). Reconstructing past seasonality is not only relevant for pure palaeoclimatological studies, but also for palaeobotany and archaeology, and for establishing a benchmark by which to compare recent changes in seasonality during the Anthropocene; recent research suggests seasonality in rainfall (e.g., Feng et al., 2013) and temperature (e.g., Santer et al., 2018) are shifting under modern climate change. This is particularly concerning because changing seasonality has had broad ecological and social implications in the past. For example, human dispersal through Asia was limited more by water availability than by temperature, and likely followed habitable corridors with favourable rainfall seasonality (Li et al., 2019; Parton et al., 2015; Taylor et al., 2018). Also, the domestication and dispersal of crops are linked to rainfall seasonality because optimal growth conditions depend on hydrological conditions. In the Fertile Crescent, barley and wheat were sown in autumn, because in this semi-arid region the winter rains are the limiting factor for their prosperity (Spengler, 2019). Similarly, abundant evidence now exists that variability in seasonal rainfall has played a key role in the waxing and waning of major civilisations (Hsiang et al., 2013; Kennett et al., 2012).

Despite the clear importance of reconstructing palaeoseasonality, it is rarely directly observable in climate proxy records. The obfuscation of seasonality by undersampling or aliasing is often a consequence of logical and pragmatic choices designed to maximise returns from available resources. Ideally, analyses would resolve nearly the full climate signal residing within every stalagmite, but this is neither logistically (given the time and funding required) nor realistically (given that the karst system transmutes the signal) possible.

Here we review both the advantages of obtaining palaeoseasonality information and methods for its reconstruction using stalagmite geochemistry and modelling, as well as common issues in extracting this information. A short review of the history of speleothem science and techniques frames the discussion and highlights how speleothems have become the premier archives for annual- to sub-annual scale terrestrial climate reconstructions, particularly during the Quaternary. We also suggest a methodology to maximise the likelihood of successfully extracting palaeoseasonality information from a stalagmite, including evaluating the hydrological characteristics of the drip feeding a stalagmite sample prior to collection, modelling palaeoseasonality from lower resolution data, and determining the seasonality of the climate at (and in regions near) the site.

## 2. Background and technique development

Very early studies demonstrated the potential of stalagmites to record climate information (Allison, 1923, 1926; Broecker, 1960; Orr, 1952). However, the real growth in the application of stalagmites as climate archives occurred after the convergence of Thermal Ionisation Mass Spectrometry (TIMS) uranium-thorium dating of stalagmites in the 1990s (e.g., Edwards et al., 1987; Edwards and Gallup, 1993) (which allowed accurate dating) and high resolution sampling techniques in the 2000s (permitting the reconstruction of climate on sub-decadal timescales). The subsequent development and proliferation of multi-collector inductively coupled plasma mass spectrometry (MC-ICP-MS) permitted extraordinarily robust (precise and accurate) chronological control (e.g., Cheng et al., 2013; Hellstrom, 2003; Hoffmann et al., 2007), while the development of a variety of microanalytical techniques provided climate proxy information of an unparalleled temporal resolution. The realisation in the late 1990s (Roberts et al., 1998) and early 2000s that stalagmite carbonate trace element compositions and isotope ratios often vary seasonally (Baldini et al., 2002; Fairchild et al., 2000; McMillan et al., 2005; Treble et al., 2003; Treble et al., 2005b) opened the door to the investigation of palaeoseasonality on an unprecedented level.

## 2.1. Increasing resolution of analysis

Immense technical progress has facilitated the transition from the first speleothem studies, which broadly placed periods of speleothem growth into the global climatic context (Harmon, 1979; Hendy and Wilson, 1968; Thompson et al., 1975), to studies adopting increasingly detailed sub-annual resolution sampling (Fairchild et al., 2001; Johnson et al., 2006; Liu et al., 2013; Mattey et al., 2008; Maupin et al., 2014; Myers et al., 2015; Ridley et al., 2015b; Ronay et al., 2019; Treble et al., 2005a). Methodological developments, particularly after the mid-2000s and particularly with respect to trace element analysis, greatly reduced the required sample size and increased measurement precision. This included the widespread adoption of micromilling techniques (Spötl and Mattey, 2006), laser ablation (Müller et al., 2009; Treble et al., 2003), secondary ionisation mass spectrometry (Baldini et al., 2002; Fairchild et al., 2001; Finch et al., 2001; Kolodny et al., 2003; Orland et al., 2008, 2009), and the development of protocols for stable carbon and oxygen isotope measurements with reduced sample sizes (Breitenbach and Bernasconi, 2011), including cold-trap methods capable of analysing less than 5 µg of carbonate powders (Vonhof et al., 2020).

Here, we apply the recently compiled Speleothem Isotope Synthesis and Analysis (SISAL) database v1b (Atsawawaranunt et al., 2018; Comas-Bru et al., 2019) to document the evolution of speleothem stable isotope record resolution. SISAL was created with the primary objective of providing access to a comprehensive repository of published stalagmite $\delta^{18}O$ records to the palaeoclimate community and for climate model evaluation (Comas-Bru and Harrison, 2019; Comas-Bru et al., 2019). SISALv1b contains 455 speleothem records (i.e., SISAL 'entities') from 211 globally distributed caves published since 1992 (Comas-Bru et al., 2019). More than half the records (264) included in the database cover at least portions of the last 10,000 years.

To investigate how stable isotope record resolution has evolved over the last three decades, we extracted all records from the database and calculated their temporal resolution as the absolute difference between two consecutive samples. Hiatuses and gaps in the individual records were excluded from the analysis, as these would have erroneously suggested much lower resolution than that actually present. In a second step, we performed the same calculation, considering only Holocene records.

The analysis reveals how the number of speleothem stable isotope records steadily increased with publication year (Figure 1), highlighting the increased popularity of speleothem science over the past three decades. A trend of increasing temporal resolution with time becomes apparent after binning all records published in the same year and calculating their mean resolution (Figure 1). This trend becomes even clearer when only Holocene records are considered, with a particularly striking increase in resolution over recent years (post-2010) (records pre-2010: mean resolution = 50.1 years, STDEV = 38.9 years; records between 2010 and 2018: mean resolution = 16.5 years, STDEV = 7.4 years), and is likely related to the widespread adoption of microanalytical advances. Additionally, a record's resolution will typically depend on the time period covered by the record; in general, resolution is higher in Holocene records compared to the full dataset, which includes older records as well. This partly arises because of greater availability of independent data and information on climate conditions during more recent time intervals, thus requiring higher resolution records to tackle relevant research questions. It may also be partially due to typically lower growth rates during the last glaciation compared to the Holocene. However, overall, only nine of the records in SISALv1b have resolution <0.5 years, directly allowing for investigations of paleoseasonality. This highlights the difficulties often encountered with conventional sampling techniques, as this compilation only includes stable isotope records, and does not consider other methods (e.g., laser ablation trace element analysis), which can generate higher resolution time-series. The increasing resolution possible via technological developments has largely involved the analysis of trace elements, whereas stable isotope analysis still predominantly relies on micromilling or drilling techniques.

**2.2. Transition from temperature to rainfall amount to seasonality**

Early speleothem palaeoclimate studies focused on using $\delta^{18}$O to generate quantitative cave temperature records (Gascoyne et al., 1980; Hendy and Wilson, 1968; Lauritzen, 1995; Lauritzen and Lundberg, 1999), based on the insight that oxygen isotope fractionation during carbonate deposition is temperature dependent (Epstein et al., 1951; O'Neil et al., 1969), and building on similar work on marine carbonates (Emiliani, 1955). It was quickly recognised however that speleothem $\delta^{18}$O is a complex mixed signal reflecting variations in cave temperature, changes in dripwater isotope composition, and various kinetic effects, which severely hamper the use of this proxy for quantitative temperature reconstructions (McDermott, 2004). The subsequent shift in how speleothem $\delta^{18}$O is interpreted led to its establishment as a proxy for past hydroclimate changes, including atmospheric circulation, regional temperature, moisture source dynamics, and amount of precipitation (Lachniet, 2009).

At the same time, the toolkit of geochemical proxies available to speleothem researchers continued to expand. In particular, trace element concentrations in speleothem carbonate emerged as tracers for numerous processes, from surface productivity to karst hydrology and transport (Borsato et al., 2007; Fairchild et al., 2001; Huang and Fairchild, 2001; Treble et al., 2005a). The combination of multiple proxies measured on the same speleothem provided a means to disentangle complexities regarding mixed signals in individual proxies and allowed a progressively deeper understanding of the archive and the associated processes in soil, karst, atmosphere, and cave. In tandem with these developments regarding the climate proxy development, monitoring of cave and local atmospheric conditions became increasingly important, as it was recognised that understanding sometimes highly localised controls on geochemical signatures is crucial for their interpretation (Genty, 2008; Mattey et al., 2008; Mattey et al., 2010; Spötl et al., 2005; Verheyden et al., 2008).

The presence of annual petrographic cyclicity within stalagmites was recognised very early on (Allison, 1926). The later identification of visible and luminescent annual banding (Baker et al., 1993; Broecker, 1960; Shopov et al., 1994) underscored that the deposition, mineralogy, and chemical composition of speleothems varied seasonally. However, the concept of seasonal shifts in climate variables (e.g., temperature, precipitation) as contributing to the net multi-annual climate signal did not gain traction until the early to mid-2000s (Wang et al., 2001). Cave monitoring revealed drip rate seasonality in Pere Noel Cave, Belgium (Genty and Deflandre, 1998), Crag Cave, Ireland (Baldini et al., 2006), and in Soreq Cave, Israel (Ayalon et al., 1998), and seasonality was discussed within the context of a speleothem-based trace element study at Grotta di Ernesto, Italy (Huang et al., 2001). Meteorological data were compared to seasonal trace element data for an Australian stalagmite (Treble et al., 2003), and the potential to use seasonal-scale geochemical data to reconstruct the East Asian Summer Monsoon (EASM) was investigated using a stalagmite from Heshang Cave, China (Johnson et al., 2006). Studies coupling cave environmental monitoring and 'farmed' carbonate precipitates were critical for clarifying the links between hydrological and cave atmosphere conditions on the chemistry of stalagmites, including at a seasonal scale (Czuppon et al., 2018; Moerman et al., 2014; Sherwin and Baldini, 2011; Tremaine et al., 2011). Drip monitoring was also key for establishing how cave hydrology attenuates seasonal and interannual rainfall variability, and was used to predict ENSO variability preservation within stalagmites (Chen and Li, 2018; Moerman et al., 2014). These studies all illustrate that a thorough understanding of annual geochemical cycles requires the development of extensive cave monitoring records, which highlight the complexities inherent in signal transfer from surface environment to the stalagmite.

## 2.3. Importance of monitoring for understanding the seasonal signal

Monitoring environmental conditions in and above a cave at a high temporal resolution greatly improves the accuracy of palaeoclimate interpretations derived from stalagmites. Linking proxy characteristics at a given site with current environmental conditions via monitoring is relevant for reconstructing past conditions. Although modern conditions may differ from ancient conditions, monitoring the cave environment clarifies processes operating at a site, including the timing and extent of ventilation and the general nature of a hydrological signal, acknowledging that some hydrological re-routing may have occurred through time for certain drip types.

Understanding a stalagmite geochemical proxy record is difficult without first understanding how that signal is transferred and altered from the external environment to the sample. Environmental changes affecting the seasonal signal fall under four main categories: ***i) Earth atmospheric, ii) meteoric precipitation***, ***iii) biological*** (e.g., soil processes), and ***iv) cave atmospheric.***

***Earth atmospheric*** processes affect the seasonality signal retained within stalagmites by influencing meteoric precipitation isotope ratios at the cave site. Possibly the most common atmospheric process is the seasonal variation in precipitation $\delta^{18}O$ induced by shifts in the temperature-dependent water vapour-meteoric precipitation fractionation factor. Other related changes in atmospheric processing include seasonal shifts in moisture source and pathway of the moisture package to the cave site, as, for example, in monsoonal settings.

***Meteoric precipitation*** variability regards the nature of the primary rainfall amount-derived seasonality signal. Here we include meteoric precipitation amount and seasonal distribution as separate from 'Earth atmospheric' processes (such as changes in moisture source), although clearly the latter affect the former. Meteoric precipitation is a fundamental control on stalagmite seasonality that is worth considering independently of other atmospheric processes. Stalagmites deposited in monsoonal climates (e.g., the East Asian Summer Monsoon, Indian Summer Monsoon, South American Monsoon, and Australian Summer Monsoon) with distinct wet and dry seasons are excellent examples of samples whose geochemistry generally (but not always) responds to hydrologic seasonality. In temperate mid-latitude settings with more evenly distributed rainfall, hydrological shifts might record less seasonal than inter-annual (e.g., ENSO) dynamics or possess a seasonal bias (see section 3.1) derived from effective infiltration dynamics.

**Biological (soil-derived)** seasonality is the least clearly defined control, and predominantly affects the trace element composition and carbon isotope ratio of cave percolation waters. However, evidence also exists that increased soil bioproductivity can affect oxygen isotope ratios by preferential uptake of water during the growing season during intervals with substantial surface vegetation (Baldini et al., 2005). Trace element transport critically depends on the biological activity and water supply, both factors that are inherently variable and not necessarily in-phase. Hydrology can affect biological seasonality, as leaching of organic matter and trace elements from freshly decomposed litter depends on excess infiltration. Soils may thus produce a wet season pulse of colloidal material (organics as well as weathering products) which contributes to an annual peak in trace element concentrations in some samples; such dynamics are highly site-specific. The evidence for this pulse is derived both from synchrotron-based stalagmite studies (e.g., Borsato et al., 2007) and daily-scale automated dripwater collection schemes (Baldini et al., 2012). Treble et al. (2003) suggest phosphorous enrichment in stalagmite carbonate stemming from seasonal infiltration pulses, and monitoring at Shihua Cave (China) revealed that organic carbon was transported during the wet season (Ban et al., 2018; Tan et al., 2006). Whether this pulse is truly independent from hydrological variability is unclear, but some evidence from dripwater monitoring in temperate Irish caves suggests that the seasonal trace element pulse is not associated with increased autumnal water throughput, but rather with seasonal vegetation die-back (Baldini et al., 2012). In monsoonal north-eastern India biologically-induced litter decomposition reaches a maximum in early summer (Ramakrishnan and Subhash, 1988), which increases element availability in the soil that can be leached during the entire wet season (Khiewtam and Ramakrishnan, 1993). Trace element transport may also hinge directly on the presence of natural organic matter in dripwater, which may link the dripwater directly to surface bioproductivity (Hartland et al., 2012; Hartland et al., 2011). Thus, biological seasonality is highly site-specific and likely variable through time; this and the complexities outlined above, underscore the importance of dripwater monitoring campaigns.

*Cave atmospheric* variability can also impart a seasonal signal to a stalagmite geochemical record. Seasonal changes in cave air mixing with outside air lead to conditions within the cave that lower cave air carbon dioxide partial pressure ($p$CO$_2$) and potentially even contribute to dripwater evaporation, promoting calcite deposition. Cave atmosphere variability, induced by ventilation (through thermal gradients or changing wind patterns) therefore affects the calcite deposition seasonality, as well as kinetic fractionation amount. Excellent examples of caves whose stalagmites are affected by this variability include New St. Michael's Cave (Gibraltar) (Mattey et al., 2016; Mattey et al., 2010) and numerous caves in Central Texas (Banner et al., 2007; Breecker et al., 2012; Cowan et al., 2013; Wong et al., 2011). These effects are discussed in detail below (Section 3).

3. Issues inherent to speleothem-based high-resolution climate reconstructions

Detecting any seasonal component in a stalagmite climate signal includes quantifying growth rate and input signal seasonality. It is worth noting that the input signal is sometimes unexpected, and a thorough site monitoring scheme can help identify the main contributing factors. For example, although many trace element ratios (and particularly Mg/Ca) are affected by recharge (often via prior calcite precipitation (PCP) mechanisms (Fairchild and Treble, 2009)), other factors can also influence (seasonal) stalagmite geochemistry. This is the case at ATM Cave, Belize, where various trace element/calcium ratios (including Mg/Ca) increase in concentration at the beginning of the annual rainy season, and are probably linked to dry deposition during the preceding dry season followed by transport to the stalagmite with the onset of the rainy season (Jamieson et al., 2015). In other cases, the advection of atmospheric aerosols directly into the cave can affect the stalagmite trace element signal (Dredge et al., 2013). Seasonal non-deposition caused by either drying of the feeder drip or by seasonally high cave air $p$CO$_2$ can bias any record where every data point integrates more than a few months of deposition. From this perspective, most stalagmite records include palaeoseasonality information to some extent, but, without appropriate monitoring strategies in place, deconvolving the extent to which the shifting seasonal signal dominates the overall record is difficult.

### 3.1. Mixing within the aquifer

The degree of recharge mixing within the aquifer and epikarst is a fundamental control on the preservation of a seasonality signal within stalagmites. A long residence time and/or thorough mixing within the overlying aquifer can greatly attenuate any hydrological seasonal signal, and understanding the hydrology feeding a cave drip is therefore critical (Atkinson, 1977; Ayalon et al., 1998; Baker et al., 1997; Baker and Brunsdon, 2003; Baker et al., 2019; Kaufman et al., 2003). For conservation and logistical reasons, monitoring and classification of the drip should ideally occur prior to sampling a stalagmite.

Smart and Friedrich (1987) undertook one of the earliest efforts to comprehensively categorise cave drips. Their scheme involved measuring drip rates at G.B. Cave, in the Mendip Hills, UK, and parameterising them by plotting maximum drip rate versus the coefficient of variation (C.V.; the standard deviation divided by the mean multiplied by 100). Baker et al. (1997) later modified the scheme, dividing drips into six categories (seepage flow, seasonal drip, percolation stream, shaft flow, vadose flow, subcutaneous flow). Other classification schemes (e.g., Arbel et al., 2010; Arbel et al., 2008) focussed on analysing drip hydrographs, and suggested terminology such as 'post-storm', 'seasonal', 'perennial', and 'overflow', which are broadly consistent with the categories introduced by Smart and Friedrich (1987). The introduction of automated drip loggers revolutionised the field (Mattey and Collister, 2008), partly by ensuring that short-lived hydrological events were not missed. This ensured a substantially more robust characterisation of drips than that possible via manually measuring drip rates only during on-site visits.

Understanding the hydrology feeding a stalagmite is fundamental for determining if a stalagmite retains a seasonal signal. Drip rate is controlled by surface processes (e.g., meteoric precipitation, evaporation, soil moisture capacity, and susceptibility to runoff) and aquifer characteristics including reservoir capacity and bedrock permeability (Markowska et al., 2015; Treble et al., 2013). Bedrock pathways recharging a drip are broadly divisible into diffuse (or 'matrix'), fracture, and conduit flows (Ayalon et al., 1998; Baker et al., 1997; Perrin et al., 2003; Smart and Friedrich, 1987), and recent models suggest that many drips are a combination of diffuse and fracture flow. Diffuse permeability typically refers to either the primary intra-granular bedrock permeability or to secondary permeability along fine fractures, and is characterised by a slow response to precipitation events and a large reservoir capacity (Atkinson, 1977; Smart and Friedrich, 1987). Fracture permeability relates to potentially solution-enlarged bedding plane partings and joints and is characterised by a rapid to intermediate response to precipitation events, and a low to moderate storage capacity. Conduit permeability refers to often solutionally-enlarged pipe-like openings >1 cm in diameter (Atkinson, 1977; Smart and Friedrich, 1987). Such conduit flow is characterised by a rapid response to storm events followed by a rapid return to baseline flow (Baldini et al., 2006), and often carries chemically aggressive waters that do not allow secondary carbonate deposition. Large conduits or bedding planes may intersect a network of more diffuse hydrological pathways, leading to dual-component flow where the fracture is itself fed by some diffuse recharge in addition to the fracture flow. The hydrologic permeability of the fracture flow component compared to the diffuse flow component essentially defines the drip type; 100% diffuse flow would exhibit no response to storm events, whereas 100% fracture flow would usually have no drip except for immediately following storm events large enough to activate the pathway (Figure 2). Most drips would fall along the spectrum between these two endmembers; a constant base drip (the diffuse flow component) combined with a variably rapid response to storm events (the fracture flow component).

From a seasonality perspective, pure fracture-flow drips vary considerably seasonally but may experience occasional dripwater undersaturation and/or drying, and consequently the resultant stalagmite could have abundant 'crypto-hiatuses' (hiatuses in growth too brief to leave a clear petrographic expression, or appear in chronological models (Stoll et al., 2015), also referred to as 'microhiatuses' (Baker et al., 2014; Moseley et al., 2015)). We suggest that if these hiatuses are demonstrably seasonally, 'seasonal hiatus' is appropriate terminology. Drips characterised by 100% diffuse flow would be stable with little hydrological or biological seasonality. Although the likelihood for seasonal hiatuses or drying is low for stalagmites fed by diffuse flow, the seasonal signal is probably muted, unless at a site where the seasonal signal is controlled by a forcing other than hydrological variability (see Section 2.4.). The optimal hydrology for imparting seasonality onto a stalagmite is a drip fed by moderately diffuse flow that is responsive to monthly-scale shifts in rainfall, but that does not have a substantial fracture component to transmit event-scale (and possibly undersaturated) water.

### 3.2. Non-deposition and seasonal bias in samples

Although growth hiatuses lasting longer than a few years are often (but not always) apparent within stalagmites as horizons of detrital material followed by competitive growth of carbonate crystals (Broughton, 1983), brief growth hiatuses occurring seasonally are often undetectable (though occasionally they have a petrographic manifestation). Thus, the existence of these seasonal hiatuses is often inferred by applying monitoring data to isolate intervals through the year where environmental conditions suggest temporary non-deposition could exist. Because drip rate is one of the fundamental controls on stalagmite growth (Genty et al., 2001), the use of drip loggers to detect seasonal drying of the stalagmite feeder drip is important for understanding whether a stalagmite record excludes a certain season's climate information.

Additionally, careful examination of sample petrography can reveal important insights into the nature of the climate signal retained by a stalagmite. Petrographic microscopy helps in identifying growth interruptions caused by lack of water, and dissolution features caused by undersaturated dripwater. An excellent example of this approach exists for Holocene stalagmites from northern Spain (Railsback et al., 2011; Railsback et al., 2017); the analysis reveals horizons of dissolution (termed Type 'E' surfaces), interpreted as reflecting occasional undersaturation of the feeder drip. Other examples of careful petrographic analysis informing seasonality studies are provided from Drotsky's Cave, Botswana, where the alternating wet and dry seasons are manifested by alternating calcite and aragonite (respectively) laminae (Railsback et al., 1994) and from Grotta di Carburangeli, Italy, where columnar fabrics were interpreted as reflected pronounced seasonal drip rate variability (Frisia, 2015).

Cave air carbon dioxide concentrations ($p$CO$_2$) are inversely linked to stalagmite growth rate (Banner et al., 2007; Sherwin and Baldini, 2011). For example, in a study of three caves across Texas, it was observed that farmed calcite growth rate was inversely correlated with cave air $p$CO$_2$ (Banner et al., 2007). Negligible calcite growth and even seasonal hiatuses occurred during the warmest summer months, when cave air $p$CO$_2$ increased due to low cave ventilation rates (Banner et al., 2007). Elevated cave air $p$CO$_2$ discourages the dripwater's thermodynamic tendency to degas CO$_2$, thereby slowing the carbonate precipitation rate. In most caves where the entrance is located above the rest of the cave, outside air with low $p$CO$_2$ advects into the cave when the outside air density becomes greater than the cave air density (e.g., Spötl et al., 2005). This is usually driven by temperature gradients; colder, denser air moves down into a cave during winter, lowering the cave air $p$CO$_2$ and encouraging stalagmite growth (James et al., 2015). However, cave air $p$CO$_2$ does not act in isolation, but instead the critical growth determining variable is the differential between cave air $p$CO$_2$ and dissolved CO$_2$ in dripwater (Baldini et al., 2008). Carbonate deposition thus could increase in the high cave air $p$CO$_2$ season if the dripwater had equilibrated with an atmosphere with even greater seasonal dissolved CO$_2$ increases (e.g., stemming from seasonal soil bioproductivity increases) which exceed those of the cave atmosphere. These types of drips are generally quite responsive to rain events, so determining if a seasonal growth bias exists should incorporate both hydrology and cave atmospheric chemistry. Drips with stable drip rates, that are not responsive to storm events may have more constant dissolved $CO_2$ and therefore seasonal deposition rates that are affected exclusively by cave air $pCO_2$ dynamics. However, several recent publications suggest that dripwater equilibrates not only with soil air, but also with a reservoir of carbon dioxide within the unsaturated zone of aquifers (termed 'ground air') that may have very high $pCO_2$ values (2 to 7%), much higher than typical soils (0.1 to 2%) (Baldini et al., 2018; Bergel et al., 2017; Markowska et al., 2019; Mattey et al., 2016; Noronha et al., 2015). Thus, it is possible that drip dissolved $CO_2$ is often near-constant, having equilibrated with a ground air reservoir of near-constant $pCO_2$, and that carbonate precipitation is anticorrelated with cave air $pCO_2$ regardless of drip type, although this requires further research. The complexities of cave atmospheres are now reasonably well understood, but more long datasets describing the dissolved $CO_2$ of cave drips are essential for determining the variability of cave percolation waters.

Although a temperate-zone (Peel et al., 2007) cave's tendency to ventilate during the winter is generally predictable from seasonality in external temperature (James et al., 2015), occasionally cave geometry provides a more dominant control. In New St. Michael's Cave in Gibraltar, ventilation is driven by seasonal changes in wind speed and direction (Mattey et al., 2016; Mattey et al., 2009). The cave experiences the lowest cave air $pCO_2$ values in summer, and consequently growth (assuming constant drip rate) is biased towards summer (Baker et al., 2014). The cave's position high within the Rock of Gibraltar contributes to strong winds and unusual seasonal ventilation, illustrating how cave position or geometry can dominate seasonal ventilation patterns. Other examples include Bunker Cave in Germany, where an essentially horizontal plan with little altitude difference between entrances produces very little seasonal variability in $p$CO$_2$ (e.g., Riechelmann et al., 2011; Riechelmann et al., 2019), and Císařská Cave (Czech Republic) where a U-shaped cave produces nonlinearities between air temperature, density, and ventilation (Faimon and Lang, 2013).

Because seasonal hiatuses can lack either a petrological or a geochemical manifestation, cave monitoring is critical for assessing the likelihood of seasonal non-deposition (Shen et al., 2013). Stalagmite growth rate modelling, informed by cave monitoring data, can provide invaluable information regarding how seasonal growth variability affects geochemical climate proxy records integrating more than one year's worth of growth. For example, seasonal non-deposition during summer due to either high evapotranspiration-induced drip cessation or elevated cave air $p$CO$_2$ might bias lower resolution records towards wintertime rainfall values (generally towards lower $\delta^{18}$O values) (e.g., James et al., 2015) at sites where drip water is not well mixed. Stoll et al. (2012) used an inverse model to illustrate that rainfall seasonality shifts relative to the cave air $p$CO$_2$ can greatly affect PCP and consequently stalagmite trace element concentrations. Baldini et al. (2008) used theoretical stalagmite growth rate equations and theory developed previously (Buhmann and Dreybrodt, 1985; Dreybrodt, 1980, 1988, 1999), coupled with monitoring information, to model stalagmite $\delta^{18}$O for various drips within Crag Cave (Ireland). The results suggest that the amount of time integrated by the analyses, the nature of the drip, and the ventilation dynamics of the cave, all strongly modulate carbonate $\delta^{18}$O signals.

These studies all highlight how characterising the surface and depositional environment is critical for interpreting the climate signal. Either seasonal hiatuses or reduced growth may bias annual- (or coarser-) scale geochemical records towards particular seasons. Additionally, it is also important to consider how regional climate shifts may have affected a sample in the past, because modern processes may not have applied throughout the record. Understanding climate signal emplacement processes within stalagmite carbonate is therefore fundamental for building robust climate records.

**3.3. A drip classification scheme to quantify seasonal responsiveness**

Existing drip classification schemes are not designed to characterise the likelihood that a sampled stalagmite retains a hydrologically induced seasonal signal. However, such knowledge is crucial if research goals include a component of seasonal climate reconstruction. Here, we introduce a new drip categorisation scheme that not only permits the identification of stalagmites most likely to retain a hydrology-modulated seasonal climate signal, but that also helps predict the general nature of the climate signal within any sample. This is important for both the accurate interpretation of stalagmite palaeoclimate records, but also for cave conservation (i.e., to maximise the usefulness of collected samples for the purpose of the research goals) and for the appropriate usage of research-related resources. A seasonal-resolution stable isotope record of any length requires considerable resources, and we hope that this new drip classification scheme will help direct these resources to appropriate stalagmite samples.

The scheme's essence is the collection of (ideally) at least one year of hourly drip rate data for a drip feeding a stalagmite of interest. For every month, the minimum and maximum hourly drip rate values are extracted. When plotted, these data reveal the extent to which the drip is affected by seasonal activation of fracture permeability, and what proportion of the drip consists of diffuse 'baseflow' (and whether this varies through the year). Drip categorisation then involves evaluating the distribution of the datapoints, and is described with terminology broadly consistent with the Smart and Friedrich (1987) scheme. Because the classification scheme uses multiple data points per site, a very large number of possible combinations of descriptors are possible. For example, some drip sites (e.g., drip site YOK-LD within Yok Balum Cave, Belize; (Ridley et al., 2015a)) are fed by a slow diffuse flow most of the year, where the minimum and maximum monthly drip rates are almost identical (Figure 3). However, during wetter months an overflow route is activated, and the maximum drip rate increases substantially, whereas the minimum remains the same; this would be characterised as a diffuse drip with a seasonally active overflow component. If this overflow component is saturated with respect to calcite or aragonite, some seasonal signal may be preserved, but if the overflow water is undersaturated a stalagmite fed by this drip type has less potential for seasonal climate reconstructions. Similarly, drip YOK-SK is characterised by almost entirely invariant diffuse recharge and would not record seasonal changes in recharge (Figure 3). At Leamington Cave (Bermuda), drip BER-drip #5 is fed by diffuse recharge during drier intervals of the year, but during wetter months more water is routed to the diffuse flow, increasing the base flow (Walczak, 2016). Consequently, the drip does experience some seasonality without risk of undersaturation, and thus a stalagmite fed by it should retain hydrology-induced seasonality.

In this new drip classification plot, drips that are expected to produce stalagmites that retain the clearest seasonal signal are those that plot with a slope approaching unity. In other words, those that are not fed by either an extremely diffuse drip or an extremely flashy drip, and that consequently respond to seasonal rainfall shifts without transient extreme rapid drip rate episodes caused by individual storm events (which may lead to dripwater undersaturation and signal loss). The two drip sites plotted in Figure 3 that best display this type of behaviour (drips YOK-G and BER-drip #5) have both yielded stalagmites retaining exceptional seasonal signals, stalagmites YOK-G (Ridley et al., 2015b) and BER-SWI-13 (Walczak, 2016). Other drip sites that have a slope approaching unity and have a pronounced difference between the highest and the lowest set of drip rates (Figure 3B) should also produce stalagmites with well-developed records of seasonality.

Importantly, this drip classification scheme equally helps to identify drips that are unlikely to produce good seasonality records. For example, stalagmites fed by drips that are invariant throughout the year would not record hydrologically-induced seasonality (although a seasonal signal might still be preserved based on non-hydrological factors – see Section 2.4). Stalagmites fed by drips that have one or more monthly values plotting at the origin (i.e., no drips for an entire month, Figure 3D) would contain seasonal hiatuses and would consequently not record that interval's climate information. Drips where the diffuse flow component (i.e., the monthly minimum flow) remains constant but the fracture flow component (i.e., the monthly maximum flow) changes considerably (Figure 3C) may experience undersaturation and either non-deposition or even corrosion of the stalagmite.

This classification scheme comes with some caveats. First, as discussed in Section 2.4., it is possible that the seasonality signal is imparted onto the stalagmite independent of hydrology. For example, if seasonal cave ventilation controls the seasonality signal, the application of the scheme would differ. At a site with strong seasonal ventilation, a stalagmite deposited by a purely diffuse flow-fed drip would reflect a largely cave atmospheric seasonality signal (i.e., with no hydrological seasonality). This would reduce the complexity of the geochemical signal and obviate the need to deconvolve hydrological- and cave atmosphere-induced seasonality from any geochemical record produced. Second, some drips are so-called 'underflow' drip sites, which respond to recharge linearly up until a maximum drip rate and then become unresponsive to further recharge increases. This is often caused by a constriction in the flow pathway leading to the water egress point into the cave. Despite the lack of variability at high flow, the dripwater is still in dynamic equilibrium with recharge (unlike high residence time diffuse flow fed sites) and the stalagmite may reflect the dripwater isotopic variability. Similarly, some drips are affected by piston flow, whereby an increase in hydrologic head might push through a slug of older water, leading to an instantaneous response to recharge but of water with a signature of 'old' water; careful monitoring can identify and mitigate these issues (see Section 3.4). Despite these caveats, this drip evaluation scheme will hopefully provide an efficient means for identifying actively growing stalagmite samples most likely to record a seasonal climate signal prior to collection of that sample.

### 3.4. Dripwater oxygen isotope seasonality

The extent that cave dripwater $\delta^{18}O$ ($\delta^{18}O_{dw}$) values reflect the $\delta^{18}O$ of meteoric precipitation ($\delta^{18}O_p$) is critical to climate studies and for understanding the palaeoseasonality signal in particular. Many publications have investigated the relationship between $\delta^{18}O_p$ and $\delta^{18}O_{dw}$ (Ayalon et al., 1998; Baker et al., 2019; Baldini et al., 2015; Bar-Matthews et al., 1996; Cruz Jr. et al., 2005; Duan et al., 2016; Feng et al., 2014; Harmon, 1979; Luo et al., 2014; Markowska et al., 2016; Mischel et al., 2015; Moquet et al., 2016; Moreno et al., 2014; Oster et al., 2012; Pu et al., 2016; Riechelmann et al., 2011; Riechelmann et al., 2017; Surić et al., 2017; Tadros et al., 2016; Tremaine et al., 2011; Verheyden et al., 2008; Wu et al., 2014; Yonge et al., 1985; Zeng et al., 2015). Depending on the drip site's hydrological characteristics (Arbel et al., 2010; Baker and Brunsdon, 2003; Smart and Friedrich, 1987), $\delta^{18}O_{dw}$ values may reflect $\delta^{18}O_p$ on timescales ranging from the annual weighted mean (Baker et al., 2019; Cabellero et al., 1996;

Chapman et al., 1992; Yonge et al., 1985) to individual (intense) recharge events (Atkinson et al., 1985; Frappier et al., 2007; Harmon, 1979).

Factors such as depth below surface, residence time and mixing of the water within the unsaturated zone, soil depth and texture, and aquifer hydraulics can vary between drip sites. Important reservoirs for storage and mixing of effective rainfall are documented as the soil and epikarst zones (Cabellero et al., 1996; Chapman et al., 1992; Gazis and Feng, 2004; Perrin et al., 2003; Yonge et al., 1985). Rainwater infiltrating into the soil reservoir is variably lost to evapotranspiration but in karst regions preferential recharge through dolines and grikes may occasionally circumvent the soil and related evapotranspiration (e.g., Hess and White, 1989). Dripwater $\delta^{18}O$ and $\delta D$ values potted relative to the local meteoric water line can detect secondary evaporation from infiltrating water (Ayalon et al., 1998; Breitenbach et al., 2015). Bar-Matthews et al. (1996) observed a 1.5 ‰ $\delta^{18}O_{dw}$ enrichment relative to rainwater and attributed this primarily to seasonal evaporation in the soil and epikarst zones above their Israeli cave site. Evaporative enrichment of infiltrating rainwater is greater in arid and semiarid regions than in temperate regions where conditions of water excess occur through much of the year (Markowska et al., 2016; McDermott, 2004). Any excess, non-evapotranspired water is then transmitted to the epikarst, karst, and finally the cave. Dripwater residence times in the aquifer or epikarst are highly variable, ranging from minutes to years, depending on soil thickness, hydraulic properties (Gazis and Feng, 2004), and drip pathway (e.g., diffuse vs. conduit flow) (Baldini et al., 2006). Mixing of infiltrating rainwater with existing epikarst water can buffer the climate signal and reduce seasonal $\delta^{18}O_{dw}$ variability from muted to invariant (within analytical error, and assuming no cave atmosphere-induced seasonality) (Baker et al., 2019; Breitenbach et al., 2019; Onac et al., 2008; Schwarz et al., 2009). At some cave sites, $\delta^{18}O_{dw}$ does not necessarily correlate with $\delta^{18}O_p$ shifts, most likely due to mixing within the aquifer (Moquet et al., 2016), underscoring that different hydrologies produce stalagmites retaining different environmental signals.

A recent global compilation of available dripwater monitoring data has further clarified the relationship between climate (e.g., mean annual temperature and annual precipitation) and $\delta^{18}O_{dw}$ (Baker et al., 2019). In cooler regions where mean annual temperature (MAT) < 10°C, $\delta^{18}O_{dw}$ most closely reflects the amount-weighted $\delta^{18}O_p$ (i.e., evaporation from the soil and epikarst does not exert much influence). In seasonal climates with MAT between 10°C and 16°C, $\delta^{18}O_{dw}$ values generally reflect the recharge-weighted $\delta^{18}O_p$ (see Fig. 1 of (Baker et al., 2019)). In regions where MAT > 16°C, $\delta^{18}O_{dw}$ is generally higher relative to amount-weighted precipitation $\delta^{18}O_p$ because fractionation processes related to evaporative effects on stored karst water are more substantial (Baker et al., 2019). Stalagmite $\delta^{18}O$ records from regions experiencing high temperatures and/or aridity will probably not reflect rainfall $\delta^{18}O$ (Baker et al., 2019).

### 3.5. The uniqueness of each stalagmite record

Recent publications have made a case for the importance of replication in stalagmite geochemical records (Wong and Breecker, 2015; Zeng et al., 2015), which is a worthwhile and useful goal. Producing the same geochemical record from multiple samples ensures that no analytical issues exist and can facilitate correlating records whose growth intervals overlap in regions and for time periods with high signal-to-noise ratios. Particularly in cases where evidence for a short-lived climate anomaly exists, replication from within the same sample and from other stalagmites is critical. However, stalagmite geochemistry is affected by a myriad of variables, and the precise combination of factors affecting any one sample are essentially unique. Thus, every stalagmite retains a different component of the environmental signal, and a lack of reproducibility does not necessarily indicate that a record is 'incorrect' or flawed. Even stalagmites that are affected by strong kinetic effects retain accurate environmental data; it is a matter of recognising this control and basing any interpretations accordingly.

Unless two stalagmites are fed by a very similar drip type (often two samples growing near each other whose feeder drips share the same hydrological pathway), stalagmite records from the same cave may not match. This is a clear consequence of the diversity of possible drip pathways feeding individual stalagmites. For example, a stalagmite growing underneath a diffuse drip fed by an extremely low hydrologic permeability pathway that is unresponsive to large rain events would not contain the same record as a stalagmite growing underneath a drip with no diffuse component but that is instead fed by fracture flow. The former (diffuse flow-fed) stalagmite may retain long-term climate information but lack seasonal-scale information, whereas the latter (fracture flow-fed) stalagmite may retain some seasonal environmental information but may also experience occasional undersaturation following large rain events, leading to hiatuses and information loss. The fracture flow-fed stalagmite may have a more rapid overall growth rate but may experience flow re-routing and stochastic drip variability due to solutional enlargement of the fracture pathway, potentially leading to a shorter overall growth interval due to the eventual diversion of water away from the stalagmite. Once cave- and site-specific ventilation factors are considered as well, it is apparent that no two stalagmites can yield precisely the same record; rather it is imperative to understand the environmental conditions recorded by each individual sample. If the goal is to reconstruct seasonality, it is important to understand the nature of the seasonality signal for each potential sample, e.g., whether the sample is affected by hydrological seasonality or cave atmospheric seasonality. In the latter case, it is then favourable to select a stalagmite from a diffuse flow drip in order to simplify the extraction of the seasonal ventilation signal.

The considerable range of stalagmite records possible, even from the same site, is potentially advantageous. The individuality of stalagmite records may yield a powerful tool for the quantitative reconstruction of historically elusive environmental variables. For example, differences in oxygen isotope ratios between two samples from the same site could reflect in-cave temperature-induced kinetic fractionation effects, and modelling (Deininger and Scholz, 2019; Deininger et al., 2016; Dreybrodt, 1988; Dreybrodt and Deininger, 2014; Riechelmann et al., 2013) could theoretically yield the cave temperature, potentially even at a seasonal resolution. This perspective is consistent with the recent appreciation that speleothems deposited at isotopic equilibrium are extremely rare (Daëron et al., 2019; Mickler et al., 2006) and that kinetic effects are an integral part of the environmental signal retained by stalagmites (Millo et al., 2017; Sade and Halevy, 2017). The concept that kinetic effects are undesirable is a vestige of early studies attempting to derive absolute palaeotemperatures from stalagmite oxygen isotope ratios, in which case kinetic effects do indeed interfere with the extraction of the desired signal. However, because stalagmite $\delta^{18}O$ values are no longer considered pure in-cave temperature proxies, kinetic effects no longer present a serious issue, provided that they are considered within any interpretations. In fact, because kinetic effects often vary in sync with the primary rainfall signal (e.g., kinetic effects tend to occur during drier periods accentuating the already elevated stalagmite $\delta^{18}O$ and $\delta^{13}C$ signature) they tend to help the climate signal stand out above background noise.

Stalagmite climate reconstructions are usually based around one record or an overlapping series of records; future research could use the differences between two records (considering in-cave kinetic effects) to reconstruct aspects of the environmental signal, including seasonal temperature shifts. Recent research utilising several stalagmites from along the same moisture trajectory across a wide region to reconstruct oxygen isotope systematics and temperature represent an exciting development in speleothem climate sciences (Deininger et al., 2017; Hu et al., 2008; McDermott et al., 2011; Wang et al., 2017), and similar methodologies could reveal in-cave fractionation processes that are ultimately relatable to temperature, potentially on a seasonal-scale. For example, changes in outside temperature-induced ventilation may affect samples fed by different hydrologies differently (promoting more kinetic fractionation in slower dripping sample), and comparing the isotope ratio records may reveal the range of external seasonal temperature variability. We suggest that the comparison of multiple coeval stalagmite geochemical records from within the same cave site is a crucial research frontier that is well worth investigating further.

## 4. Analytical techniques

Direct detection of seasonal variations in stalagmite geochemical parameters requires sampling or analysis at sufficiently high spatial resolution to mitigate signal averaging (Figure 4). Sampling frequency should approach monthly resolution to detect a seasonality signal and to avoid aliasing issues during intervals with slower growth. This necessitates careful consideration prior to analysis to ensure both sufficient sampling resolution to detect seasonal-scale variability, and sufficient material for the analytical method. In addition to the pre-analysis considerations, we also recommend publishing complete micro-analytical data tables, in order to increase transparency. Below we discuss common microanalytical techniques capable of palaeoseasonality reconstruction and compare advantages and disadvantages of each.

## 4.1. Sampling for palaeoseasonality

Sub-sampling stalagmites for geochemical analysis requires careful planning and execution. We recommend a thorough reconnaissance of a sample's petrography using microscopy prior to geochemical analysis. The conversion of a sample into polished thin sections can provide critical information but is destructive. Reflected light microscopy provides a non-destructive alternative that can yield crucial information regarding crystal growth habit, the location of possible hiatuses, inclusions, and porosity.

The various methods available for the extraction of proxy data all require different sample amounts depending on analytical limits of detection and other factors (Fairchild et al., 2006). Methods are broadly categorizable as destructive and non-destructive, depending on the amount of material required. The former is further divisible into: i) macro-destructive (e.g., cuttings for fluid inclusion studies, low-concentration proxies like biomarkers or DNA) (e.g., Blyth et al., 2011; Vonhof et al., 2006; Wang et al., 2019a), ii) meso-destructive (e.g., conventional and micro-milling for U-series samples, stable isotopes, ICP-OES, $^{14}$C) (e.g., Lechleitner et al., 2016a; Ridley et al., 2015b; Spötl and Mattey, 2006), and iii) micro-destructive (e.g., laser ablation or secondary ionization mass spectrometer (SIMS) analyses for traditional and non-traditional isotope systems, element concentrations or ratios) (Baldini et al., 2002; Luetscher et al., 2015; Treble et al., 2007; Webb et al., 2014; Welte et al., 2016). Non-destructive methods include (but are not restricted to): i) simple desktop scanning and photography, ii) µXRF line scanning and mapping (e.g., Breitenbach et al., 2019; Scroxton et al., 2018), iii) synchrotron analyses (e.g., Frisia et al., 2005; Vanghi et al., 2019; Wang et al., 2019b; Wynn et al., 2014), iv) phosphor mapping via beta-scanning (e.g., Cole et al., 2003), v) reflected light, and fluorescence, including confocal laser fluorescent microscopy (CLFM) (e.g., Orland et al., 2012) and other microscopy techniques (e.g. SEM, EMPA, RAMAN), or vi) X-ray Computed Tomography (CT) scanning (e.g., Walczak et al., 2015; Wortham et al., 2019). The choice of technique should consider suitability for answering the targeted research questions, and logistical considerations such as sample sectioning. Although the list above categorises techniques based on their destructiveness, it does not account for sample preparation; for example, SIMS analysis uses only a small amount of sample (i.e., essentially non-destructive), but requires sectioning of the stalagmite into centimetre-scale cubes, polishing and epoxy-mounting. Another major consideration is the length of the record required; it is possible (though labour-intensive) to produce seasonal-scale records extending hundreds or even thousands of years using micromilling, but this is not practical using SIMS, unless automated protocols allowing for unattended analysis can be developed.

Although macro-destructive sampling can inform interpretations based on higher resolution data, it cannot generally reconstruct seasonality on its own. Thus, here we discuss only selected meso-, micro-, and non-destructive techniques. The focus is first on 'conventional drilling' and 'micromilling' of powder samples, which probably are the most widely used techniques to obtain material for inorganic chemistry, followed by the highly versatile, fast, and cost-effective laser ablation sampling (LA-ICPMS). SIMS requires substantial sample preparation, offers excellent resolution and is a good choice in situations requiring in-depth characterisation of a short interval. Synchrotron-µXRF (SR-µXRF) has advanced considerably over the past decade, and it is now possible to obtain high-resolution (0.5-5 µm) quantitative trace element data non-destructively through fast scanning of large samples (Borsato et al., 2019). Below we describe the relevance and applicability of these techniques towards the reconstruction of palaeoseasonality.

### 4.1.1. Conventional drilling

Conventional drilling (or 'spot-sampling') (Fairchild et al., 2006) is the drilling of powders from discrete spots that are normally separated by unsampled material, and is still amongst the most widely used methods to obtain carbonate powders from speleothems. This method is comparably fast and, with a sufficiently small drill bit (typical Ø ca. 0.2-1 mm), can achieve a spatial resolution of up to 0.3-0.5 mm along the growth axis, although more frequently the resolution is ~1 mm. Conventional drilling is ideally performed with instruments that allow computer-aided control of x-y-z dimensions, such as Sherline® or Mercantek® instruments.

With typical stalagmite growth rates of 0.1 to 0.2 mm year$^{-1}$, this technique is usually inadequate when targeting sub-annual resolution (Figure 5). If used on samples with growth rates approaching twice the sampling interval, aliasing may occur and unfavourably affect the recovery of high-frequency variability (Fairchild et al., 2006). Furthermore, this type of spot sampling usually does not integrate all the carbonate material, i.e. the time slices at the top and bottom of the hole are under-represented in the average for the drill-hole; this undersampling could miss short-lived climate excursions. Consequently, we cannot recommend conventional drilling for recovering a seasonal signal, although the technique is effective at quickly producing a lower-resolution record and is well suited for longer records of climate (e.g., those covering multiple glacial cycles), and for screening potential target stalagmites. Additionally, conventional drilling is possible on a large stalagmite slab, obviating the need for sectioning into multiple smaller slabs. A related technique which is preferred for sampling at seasonal scale is micromilling, discussed below.

### 4.1.2. Micromilling

Micromilling refers to continuous sample cutting along a trench parallel to a stalagmite's growth axis (Fairchild et al., 2006; Frappier et al., 2002; Spötl and Mattey, 2006). Usually performed with computer-controlled milling devices (such as the ESI/New Wave micromill) this technique can achieve ~10-micron spatial resolution e.g. Ridley et al., 2015b, but is critically dependent on the textural characteristics of the sample. Dense columnar, fascicular, radiaxial, or radial fibrous calcites are the most suitable material, but needle-like aragonite can also be sampled, although gaps between needle-shaped crystals may lead to loss of sample and require painstaking cleaning procedures. The sample morphology throughout the stalagmite also warrants consideration. Planar, parallel, and laterally continuous laminae across the sample are ideal, but often stalagmite laminae appear curved in a slabbed sample. These are normally convex, but in some cases are concave (particularly in the case of a 'splash' cup), and with laminae that thin towards the edges. The greater such curvature, the narrower the micromilling trough required for sub-annual (seasonal-scale) sampling (Figure 5), because a wider trench would integrate material from other laminae. Similarly, the sample should allow 2-3 mm sampling into the depth of the sample slab, and ideally the growth layers should not taper out in the third dimension. X-ray and Neutron CT scans can help visualise the 3D internal structure of the sample (Walczak et al., 2015; Wortham et al., 2019), and the appropriate milling depth.

The determination of the x, y, and z dimensions of the sampling increment is the first step of any sampling strategy (Figure 5). For seasonal resolution, this strategy will ideally permit a very small y-axis increment (the y-axis is parallel to the stalagmite growth axis). The other dimensions must then allow the collection of enough carbonate for analysis (typically 50-120 μg for carbon and oxygen stable isotopes). Depending on sample characteristics and desired resolution, dimensions of y = 10-100 μm and x = 10-300 * y μm (parallel to growth layers on the slab) are ideal (Figure 5). The sampling depth (z-axis) is best minimised because lamina behaviour into the sample is often unknown, unless CT scans of the sample exist. Larger sample masses are occasionally needed for non-traditional proxies.

A common issue in the speleothem sciences is the precise correlation between two datasets obtained via different means, for example a micromilled stable isotope dataset and a LA-ICPMS derived trace element dataset. Annual- to decadal-scale correlations are usually possible, but rarely are the records correlative on the seasonal- or even annual-scale. Comparisons are achievable using very careful measurements from a datum (often the stalagmite top), with or without the use of banding as 'landmarks' (e.g., (Johnson et al., 2006; Treble et al., 2005a)). A recent technological advance is the development of software, such as the open-source GIS-based QGIS software (Linzmeier et al., 2018), which integrates micro-imaging and analysis into a single spatial reference frame. This approach is particularly useful for organising different analyses derived from differently sectioned portions of samples and has been successfully applied to stalagmite data (Orland et al., 2019).

The problem of correlating different types of data is to some extent avoidable by sampling sufficient material with the micromill for both stable isotope and trace elemental analysis via ICP-MS. The sampled powder is divided into two aliquots, one for each analytical technique. The resultant trace element and stable isotope data permit zero-lag cross-correlations and highly robust interpretations of different environmental processes (e.g., Jamieson et al., 2016).

For example, if planned multi-proxy analyses require 0.8 mg of carbonate powder (e.g., stable isotope ratios, $^{14}$C, and trace elements), and a 50 μm spatial resolution is desired using a milling bit diameter of 0.8 mm, a 0.05 mm x 4.15 mm x 1 mm trench would suffice (assuming calcite density of 2.7 g/cm$^3$ and no sample loss via incomplete recovery); sample loss and a particularly low-density sample would require a larger volume. An often-overlooked additional consideration involves the corners that are initially unsampled when milling trenches (red corner areas, Figure 5). Depending on the drill bit diameter and trench dimensions, the corners at each end of the trench would lead to unwanted integration of material from several sample increments and thus time slices. Use of a smaller milling bit diameter minimizes this effect. Additionally, a 50% reduction of this sampling effect is achieved if a trench is milled along the growth axis prior to the high-resolution milling, or if the milled trench is adjacent to a longitudinal cut (Figure 5). Material from the first trench can be used for reconnaissance studies. Another approach yielding similar results involves collecting the desired powder, and then moving the milling bit along the horizontal sampling track (i.e., parallel to the growth layer) for a distance corresponding to half the width of the milling bit. This powder is then discarded (or collected as auxiliary powder), and the milling bit returns to the original position, ready to produce the next aliquot of powder. Either of these sampling approaches effectively reduce spatial integration of sample (Kennett et al., 2012; Myers et al., 2015; Ridley et al., 2015b), thereby increasing the likelihood of obtaining a clear seasonal signal (Figure 6). These considerations are important because many stalagmites, particularly from non-tropical localities, may have low growth rates (Railsback, 2018) that require a very high sampling resolution with minimal integration across samples to extract a seasonal signal.

Many samples may deviate from an idealised geometry, and may contain imperfections along preferred micromilling tracks, growth rate changes, or growth axis shifts. These instances may require special consideration and sample-specific solutions, such as moving to a different track within the sample or changing the resolution of the analyses in response to major changes in growth rate. In the case of the latter, interpretations should consider how changes in sampling resolution might have affected the amplitude of any seasonal cycle.

Other issues include growth layers that slope inward rather than geometrically perfect layers (where the layering is perpendicular to section) and the use of tapered rather than cyclindrical drill bits, which samples less carbonate at depth than the step size implies, and then integrates this carbonate into subsequent samples. A study comparing micromilling/IRMS and SIMS techniques on annually layered otoliths found that an offset existed between the two techniques (with SIMS yielding values ~0.5‰ lower) and that the amplitude of annual oxygen isotope signal derived via micromilling was approximately half of the SIMS signal; both of these observations are potentially explained by deviations from an ideal sample geometry, and consequently greater integration of unwanted material arising from micromilling (Helser et al., 2018). Despite these differences, both techniques were able to detect annual isotope ratio cycles (Helser et al., 2018). A thorough reconnaissance of the sample using CT scanning or other means to characterise its geometry in advance of slabbing can minimise these issues.

Other minor issues include the possible conversion of aragonite to calcite during milling, which would result in a decrease in $\delta^{18}O$ values of 0.02‰ for every 1% aragonite converted to calcite (Waite and Swart, 2015). This effect may have implications for modelling oxygen isotope variability or calculating deviations from equilibrium deposition. However, using a slower rotation rate of the milling bit (500-800 rpm) will minimise, or even eliminate, this effect. A final recommendation is to run micromilled samples through the IRMS non-sequentially (i.e., out of stratigraphic order). Ideally the laboratory environment is static and will not affect results, but any unaccounted for changes (e.g., lab temperature) may affect the analyses in a cyclical way. Running samples non-sequentially both helps ensure that any cycles detected (e.g., a seasonal cycle) are not analytical artefacts and helps to identify issues, if they exist (e.g., a persistent cycle when samples are arranged in the order that they were run).

### 4.1.3. LA-ICPMS

Laser Ablation Inductively Coupled Plasma Mass Spectrometry (LA-ICPMS) is a beam method sampling technique. A polished speleothem slab is analysed by ablating small portions of material using a laser within a sample cell. The laser (typically an ArF excimer laser at a 193 nm wavelength) physically ablates the sample, aerosolising the material which is then carried into the ICP-MS system by a carrier gas (typically helium and/or argon, with helium yielding a greater signal intensity (Luo et al., 2018)) where trace element concentrations are measured and quantified against standards of known compositions. The specific mass spectrometer set-up depends on the research question; for example, by using a quadrupole ICP-MS for elemental measurements using a reference isotope, or a multi-collector ICP-MS for isotope ratio analyses. Additional analytical set-ups are compatible with LA-ICPMS, including reaction cells, triple-quadrupoles, and split-stream analysis using two mass spectrometers in tandem (Frick et al., 2016; Kylander-Clark et al., 2013; Woodhead et al., 2016).

The advantages of LA-ICPMS for speleothem trace element analysis are numerous and include excellent spatial resolution (down to ~3 microns (Müller and Fietzke, 2016) using a rectangular aperture with long axis oriented along laminae) whilst preserving low detection limits (Figure 6). Although historically LA-ICPMS instruments used round 'spots', some laser ablation instruments are now fitted with rectangular masks (apertures), resulting in rectangular spots optimised for speleothem analysis, where the ablation spot's long dimension is oriented perpendicular to speleothem growth axis, along the x-axis (Müller et al., 2009). This permits the ablation of a surface area equivalent to large circular spot sizes, while retaining high spatial resolution in the growth direction (similar to the micromill sampling described in 4.1.2). The speed of analysis via this method is also exceptionally high, with typical scan speed of 10 μm s$^{-1}$ (e.g., (Jamieson et al., 2015)). Two-volume laser cells are now available, minimising sample damage incurred via sectioning and ensuring consistent aerosol flow within the cell. The coupling of a laser ablation system with a large-capacity gas exchange device even allows analysis under atmospheric air (Tabersky et al., 2013) although with somewhat elevated limits of detection. This technique is particularly suitable for large stalagmites, or archaeological samples, because it minimises physical sample destruction by requiring less sectioning.

The presence of a localised impurity can produce a trace elemental concentration peak even in the absence of a laterally contiguous geochemical horizon with that geochemistry. LA-ICPMS can produce elemental maps that can verify the spatial continuity of geochemical laminae of interest, particularly when combined with a square aperture (Evans and Müller, 2013; Rittner and Muller, 2012; Treble et al., 2005b; Woodhead et al., 2007). This permits the resolution of spatial relationships with greater confidence and can corroborate interpretations based on stacked and parallel line scans, thereby avoiding issues related to the overinterpretation of a small number of points. Other microanalytical techniques (e.g., SIMS, synchrotron, µXRF, etc.) can also produce elemental maps, but LA-ICPMS techniques can provide greater spatial coverage more rapidly.

The most significant disadvantage to LA-ICPMS is related to difficulties with standardisation. The use of matrix matched standards (i.e., made of the same material as the sample) during laser ablation analysis is ideal, but the limited availability, variable degrees of standard homogeneity, and accurate standardisation of carbonate materials are ongoing challenges. Orland et al. (2014) and later Müller et al. (2015) provide promising tests for a carbonate standard, albeit for a limited range of elements. Many analyses are standardised with somewhat greater uncertainty than is ideal using glasses such as NIST 620 or 622. These analyses are often regarded as semi-quantitative, with high levels of confidence regarding variability and data trends but uncertainty regarding absolute values. Another minor disadvantage is lack of precise knowledge regarding the position of individual analytical spots. The sheer number of analyses possible via this technique (often >10,000) and indistinct, continuous track means that the exact position of any one individual spot is often difficult to determine precisely, complicating the correlation with other climate proxies. This disadvantage is mitigatable by precise notetaking, syn-analytical microscopy recording, careful reflected light imaging, cross-correlation, application of QGIS or similar software, and judicious 'wiggle-matching' with other proxy records, as well as creating marker laser lines at certain intervals to further help to constrain spatial uncertainties.

## 4.1.4. Secondary ionisation mass spectrometry

Secondary ionisation mass spectrometry (SIMS) uses a primary beam of positive (often caesium) or negative (often oxygen) ions to impact a sample surface under a vacuum, 'sputtering' secondary ions into a mass spectrometer (Wiedenbeck et al., 2012). The sputtered secondary ions are then accelerated into a double-focusing mass spectrometer and counted by ion detectors (electron multiplier or Faraday cup). This analytical technique can yield both trace element analysis and stable isotope ratio data in speleothem carbonate at the micron scale, with very little damage to the sample, and with very high sensitivity (Figure 6).

The spatial resolution typically ranges between 1 to 10 µm spot size and 1-2 µm spot depth for trace elements, with stable isotope analyses historically restricted to 20–30 µm resolution (Fairchild and Baker, 2012) but now capable of achieving 10 µm resolution (Orland et al., 2019). This represents a very high-resolution method for stable isotope analysis within speleothem carbonate and is therefore ideal for detecting palaeoseasonality (Fairchild et al., 2006). The analytical resolution for trace elements is lower than when using synchrotron radiation, but with the added advantage of full quantification of concentration data and the ability to cover much greater areas of sample. Matrix matched materials, typically calcium carbonate, are used for standardisation to ensure consistent ionisation of chemical species and ablation rates (Fairchild and Treble, 2009).

Early studies of SIMS-derived trace element trends in speleothems helped to demonstrate that many stalagmites retained a seasonal signal (Baldini et al., 2002; Finch et al., 2001;

Roberts et al., 1998), representing a considerable shift in resolving power compared to the former decadal- to centennial-scale of analysis previously possible. The presence of annual trace element cycles was quickly established as the norm rather than the exception for shallow cave sites, even in the absence of visible speleothem laminations (Fairchild et al., 2001). Divalent alkaline earth metals such as magnesium and barium were suggested as palaeohydrological proxies, phosphorus as indicative of bioproductivity, and strontium as reflecting calcite growth rate and/or PCP (Fairchild et al., 2001; Fairchild et al., 2000; Treble et al., 2003). However, the need for better empirical transfer functions between speleothems and external climatic processes, and partitioning between drip waters and speleothem calcite, complicated interpretations (Fairchild et al., 2001). Subsequent process-based studies have revealed the complexity involved in interpreting trace elements at seasonal scales, highlighting the role they play in complexation with organic matter as colloids (Borsato et al., 2007), in speleothem diagenesis (Martin-Garcia et al., 2014), and the complex controls on transfer through vegetation/soil/epikarst (Hartland et al., 2009; Hartland et al., 2012), as well as controls on partitioning via internal cave microclimate and crystallographic structures (Fairchild and Treble, 2009). The use of trace element cycles obtained via SIMS as chronological markers is exemplified through the work of Smith et al. (2009), where the ability of trace element cycles to provide relative age constraints at a finer spatial resolution than traditional U-series age models is unambiguously demonstrated.

A frontier for SIMS trace element measurements lies in the potential of combining these trace element records with stable isotope measurements undertaken at sub-annual scale. Prior to the advent of SIMS techniques for stable isotope analysis, there were very few combined trace element – stable isotope studies due to the incompatibility of analytical resolution between the two parameters (Orland et al., 2014). However, the analysis of stable isotopes by SIMS now achieves a spatial resolution capable of allowing direct comparability between both isotopic and trace element indicators of seasonality (Orland et al., 2014).

SIMS stable isotope studies have investigated the $\delta^{18}O$, $\delta^{13}C$ and $\delta^{34}S$-$SO_4$ dynamics in stalagmite records (typical uncertainties (2$\sigma$): $\delta^{18}O$ = 0.2‰ (Orland et al., 2019); $\delta^{13}C$ = 0.6-0.7‰ (Oerter et al., 2016; Sliwinski et al., 2015); $\delta^{34}S$ = 1.6‰ (1$\sigma$) at 70 ppm S concentrations (Wynn et al., 2010)). Whereas each of these isotope ratios reflects changing surface environmental conditions over inter-annual timescales, only the $\delta^{18}O$ measurements by SIMS can produce records of intra-annual seasonality. Analysis of $\delta^{13}C$ in speleothem carbonate cannot be undertaken simultaneously with $\delta^{18}O$, and any available records in the literature (e.g., (Pacton et al., 2013)) are not undertaken at seasonal resolution. The apparent lack of seasonal change in cave dripwater $\delta^{34}S$-$SO_4$ (Borsato et al., 2015) has also so far prevented SIMS speleothem sulphur isotope measurements at the seasonal scale (Wynn et al., 2010). Treble et al. (2005a) produced the first $\delta^{18}O$ record unambiguously linking seasonal cycles in speleothem oxygen isotopes to rainfall dynamics and corroborated these interpretations with trace element cycles and contemporary rainfall monitoring. Subsequent work at Soreq Cave (Israel), further developed the technique to detect seasonality and links with rainfall dynamics across a range of time periods (Orland et al., 2012; Orland et al., 2009; Orland et al., 2014). Coupled annual variability in fluorescence and $\delta^{18}O$ provided a seasonal marker of annual variability in rainfall from before the climate instrumental record (Orland et al., 2012; Orland et al., 2009). Careful correlation between fluorescent banding, $\delta^{18}O$ and trace element measurements, and surface environmental conditions demonstrated that the fluorescent banding represented seasonal organic colloid flux variability into the cave.

Despite the clear advantages of utilising SIMS stable isotope analyses of speleothem carbonate to reveal seasonal patterns of rainfall delivery and drivers of climatic change, the technique also comes with its analytical challenges, including the considerable impact of geometric imperfections (e.g., sample topography, porosity, inclusions, cracks, etc) (Kita et al., 2011; Liu et al., 2015; Pacton et al., 2013; Treble et al., 2005a). In most instances, the ability to control the precise location of SIMS analyses enable geometric imperfections to be avoided, provided that i) good surface mapping can be used to identify optimal locations for analysis and that ii) post-processing can visualise geometric imperfections in each analysis pit (Orland et al., 2009). This contrasts with micromilling, where large swathes of sample are often bulked together regardless of sample porosity or imperfections. The need to use matrix matched standard materials presents similar problems of availability and homogeneity for the accuracy of data analysis as encountered with LA-ICPMS. However, recent improvements in this area, alongside improvements in sample preparation techniques have been substantial enough to enable accurate correction for instrumental drift (Valley and Kita, 2009). The impact of trace element content on carbonate $\delta^{18}O$ and $\delta^{13}C$ analyses also requires careful consideration (Sliwinski et al., 2017), but can be corrected following careful standardisation and is generally not a problem encountered through speleothem analysis where the trace element content is typically less than 1 weight %. An emerging analytical frontier concerns the impact of water and/or organic content on SIMS carbonate $\delta^{18}O$ and $\delta^{13}C$, requiring careful pre-screening of sample material and simultaneous analysis of OH- and CH- respectively (Orland et al., 2015; Orland et al., 2019; Orland, 2013; Wycech et al., 2018).

Despite these issues, SIMS remains an appealing choice for palaeoseasonality reconstruction using stalagmites due to its sensitivity and resolution. SIMS has produced some of the highest resolution records of palaeoseasonality available and will continue to play an important role in linking stalagmite records to seasonal changes in environmental conditions, particularly across discrete, short-lived events. Although the technique is not suitable for building long records, the comparison of discrete timeslices permits seasonality to be contrasted for key intervals (Orland et al., 2012; Orland et al., 2015; Orland et al., 2019).

**4.1.5. Synchrotron**

The application of Synchrotron Radiation micro X-Ray Fluorescence (SR-µXRF) to the study of speleothem carbonate opened up new possibilities in terms of greater resolving power for geochemical analysis (Kuczumow et al., 2003; Kuczumow et al., 2001). Based on the emission of electromagnetic radiation from charged electrons accelerated in an orbit, synchrotron radiation generates secondary radiation from speleothem carbonate based on the characteristic fluorescent properties of chemical elements. The excellent spatial resolution of analysis (0.5–5 microns), low detection limits, low background, and the ability to quantitatively map trace element variability across a given area has enabled the study of speleothem geochemical structures at the sub-annual timescale and in two dimensions (Figure 6). The use of XANES (X-Ray Absorption Near Edge structure) can define the oxidation state of the element under consideration, thereby adding further resolving power to determine environmental processes.

Applications range from using SR-µXRF to determine long-term (100 year) secular changes in elemental signals (Frisia et al., 2005), high resolution event imaging across sub-annual to multi-annual timescales (Badertscher et al., 2014; Frisia et al., 2008; Vanghi et al., 2019; Wang et al., 2019b), and for investigating petrological controls on geochemical composition (Frisia et al., 2018; Ortega et al., 2005; Vanghi et al., 2019). However, it is at the seasonal scale of analysis where the resolving power of synchrotron radiation has really pushed the boundaries of speleothem science.

No conventional dating technique provides an absolute timeframe at the sub-annual scale of speleothem carbonate deposition. However, linking the seasonality of external environmental processes to speleothem petrology and geochemical characteristics can yield a monthly scale resolution of trace element content. SR-µXRF was used to determine the coincidence of trace element distributions and physical calcite characteristics within annual stalagmite laminations (Borsato et al., 2007). Based on the annually laminated stalagmite ER78 from Ernesto Cave, Italy, a suite of trace elements (P, Cu, Zn, Br, Y, and Pb) were found to form an annual peak, coincident with a characteristic thin (0.5-4 µm) brown UV-fluorescent layer in each annual couplet. The brown colouration of each UV-fluorescent layer is probably due to organic acids derived from high rates of water infiltration during each autumn (Frisia et al., 2000; Huang et al., 2001; Orland et al., 2014). The transport of trace elements is associated with colloidal organic molecules (Hartland et al., 2010; Hartland et al., 2012), and leads to the incorporation of this distinctive elemental suite on a seasonal basis associated with the autumnal rains (the 'autumnal pulse' as described in Section 2.4). SR-µXRF permits the detection of variability inherent to each individual year, which then can be contrasted against the symmetrical mean annual profile. Any differences (e.g., double peaks or shoulder peaks) provide an indication that the rainfall distribution throughout that year deviated from the mean annual profile. Strontium was observed to vary inversely to colloidally transported elements (Borsato et al., 2007), possibly due to competition for binding to defect sites, thus limiting incorporation into the calcite lattice. SR-µXRF revealed seasonal patterns of zinc, lead, phosphorus, and strontium within speleothem Obi84 from Obir Cave, Austria, whose concentration peaks also coincided with the dark coloured visible laminae. These were similarly interpreted as hydrological event markers associated with autumnal infiltration but could also result from dry deposition of aerosols (Dredge et al., 2013).

SR-μXRF 2D mapping within speleothem Obi84 over three annual cycles demonstrated the effects of several infiltration events each year, present as short-lived peaks in Zn concentration and which build in magnitude towards the main autumnal flush (Wynn et al., 2014) (Figure 6). Using these event peaks as markers of autumnal flushing permitted attribution of annual sulphate cycles to summer high and winter low concentrations. At the Obir Cave site, these seasonal shifts in speleothem sulphate content were attributed to temperature-driven cave ventilation and associated cave air $p$CO$_2$ variability which controlled the dripwater pH and the sulphate:carbonate ratio. Wynn et al. (2018) later verified this proposed seasonal mechanism using controlled laboratory experiments, thereby permitting the extraction of seasonal temperature information based on the annual sulphate cycle's topology. SR-μXRF can thus extract geochemical expressions of seasonality, and the technique is well-suited to investigating changing rainfall and temperature seasonality dynamics back through time.

### 4.1.6. Data analysis

Following the geochemical analyses and data processing, the information must be interpreted. For techniques producing tens to hundreds of data points, this is not particularly challenging. On the other hand, techniques such as LA-ICPMS can produce tens of thousands of data points for multiple elements and can greatly increase the processing time on common spreadsheet programmes. To circumvent these issues, it is possible to simplify the data using a Principal Component Analysis (PCA), a multivariate statistical analysis technique which extracts modes of variation from large multivariate timeseries datasets that best describe overall variability of those datasets. The technique is ideal for large multivariate stalagmite-derived LA-ICPMS datasets (Borsato et al., 2007; Jamieson et al., 2015; Orland et al., 2014; Wassenburg et al., 2012). PCA has also been used to extract a seasonal signal from trace elemental concentrations even in the absence of visible laminae and applied towards the development of a chronology (Ban et al., 2018).

Comparing the intra-annual amplitude of a geochemical signal (Orland et al., 2012; Orland et al., 2009; Orland et al., 2014; Orland et al., 2019) from monthly-resolved datasets is ideal for extracting seasonal information from an otherwise difficult to interpret dataset. For example, Ridley et al. (2015b) used the well-developed annual carbon isotope cycles with their Belizean stalagmite to extract seasonal amplitudes, which were then interpreted in terms of the strength of the seasonal ITCZ incursion into southern Belize. Orland et al. (2015) used the topology of oxygen isotope variability within individual growth bands in a Chinese stalagmite to clarify the origin the oxygen isotope variability. Spectral analysis of well-dated samples can also reduce data complexity (Myers et al., 2015; Ronay et al., 2019). For example, Asmerom et al. (2020) used a wavelet analysis to reconstruct the strength of the wet season in Central America over the last two millennia, and to show that modern seasonality in rainfall was only emplaced in the 15th Century. Extracting a meaningful metric from numerous more complex data using statistical techniques is one way of simplifying a complex geochemical dataset.

## 5. Modelling techniques

There have been many efforts at modelling both the hydrology feeding a stalagmite and the climate signal within. Proxy system models (PSMs) describe how geological or chemical archives are imprinted with a climate signal (Evans et al., 2013). In terms of stalagmite-specific models, several exciting geochemical models now exist which can explore the emplacement of a geochemical signal in a stalagmite (Wong and Breecker, 2015), often based on established processes which govern stalagmite precipitation (e.g., (Buhmann and Dreybrodt, 1985)). Two recent examples (specifically of disequilibrium isotope fractionation processes proxy system models) are the IsoCave model, which can examine disequilibrium isotope effects in speleothems and related implications for speleothem isotope thermometry (Guo and Zhou, 2019), and the ISOLUTION model which similarly helps to better understand the effect of these disequilibrium isotope fractionation processes on stalagmite proxy records (Deininger and Scholz, 2019). The I-STAL model allows the simulation of PCP and how this affects dripwater Mg, Sr, and Ba (Stoll et al., 2012). Numerous models looking specifically at drip hydrology now exist (e.g., KarstHydroModel (Baker and Bradley, 2010; Treble et al., 2003)), and these are extremely useful for understanding how the rainfall input signal is transformed before reaching the stalagmite. Rather than using hydrological or geochemical modelling, a recent publication introduced a Monte Carlo approach to model rainfall and temperature seasonality in a stalagmite from La Garma Cave, northern Spain, over the Holocene (Baldini et al., 2019). Here, we build a second generation of this model and compare results to both synthetic and real-world input data. Whereas the older version of the model could only run a limited number of simulations and a run stopped once the model converged upon a solution (though it could be run multiple times), this next generation model is able to run a large number (user-defined; we used 1,000 simulations in the runs presented here) of simulations and retain the output of each one, permitting the creation of probability distributions for each timeslice.

This new model requires some widely available types of input data, including: i) a stalagmite-based $\delta^{18}$O record, ii) a record of regional mean annual temperature (MAT) of any resolution (e.g., borehole, marine sediments, stalagmite fluid inclusions) over the interval of interest, iii) monthly-scale modern instrumental records of rainfall and temperature above the site (or as close as possible to the site), and iv) cave air temperature and its relationship with above ground temperature. The relationship between meteoric precipitation $\delta^{18}$O and temperature at the site is useful but not required information because regional or global meteoric precipitation $\delta^{18}$O and temperature equations can provide a suitable alternative.

Essentially, the model assumes that the MAT of the cave site is similar to the MAT of the regional surface temperature input record (ii above) and produces a sine function around this value of an amplitude reflecting modern surface temperature seasonality but with random variability added to the absolute minimum and maximum temperatures (the amount of randomness is user-defined). A second sine function reflects the rainfall seasonality, and whereas the temperature wave's polarity is fixed (i.e., summers are always warmer than winters), the rainfall seasonality sine wave's polarity is allowed to flip randomly (but where only outputs that 'converge' are retained, and unrealistic results are rejected – see below). The seasonal extreme values ('extreme' meaning minima and maxima) associated with either sine function are fixed to the same calendar months, linked to the timing of the modern minima and maxima.

These two sine waves produce synthetic monthly temperature and rainfall values, which are then converted to $\delta^{18}O_p$ based ideally on local temperature-rainfall $\delta^{18}O$ relationships, or in cases where this relationship is not known, to more global equations (e.g., (Schubert and Jahren, 2015)). It is assumed that the $\delta^{18}O_p$ is conveyed to the dripwater (see discussion regarding evapotranspiration, Section 4.3) and that this is converted to carbonate $\delta^{18}O$ using the Tremaine equation (Tremaine et al., 2011) at ambient cave air temperature adjusted according to observed relationships between outside and inside air. This equation was chosen as most appropriate because its empirical nature accounts for in-cave disequilibrium fractionation processes more completely than other equations. The model therefore considers seasonal changes in rainfall but is independent of total annual rainfall. The annual amount-weighted mean modelled carbonate $\delta^{18}O$ value is then compared with the actual measured carbonate $\delta^{18}O$ value, and if it is within a certain user-defined value, it is logged as a successful simulation. If the difference between the modelled and actual carbonate $\delta^{18}O$ is greater than this value (generally ~0.1 per mil), the simulation is logged as unsuccessful. 1,000 of these coupled temperature and rainfall simulations are conducted per time slice, all the successful and unsuccessful simulations are logged, and the mean monthly modelled rainfall and temperature values calculated from the successful simulations. For a table describing the steps in the modelling process, please see Baldini et al. (2019).

**5.1 Test Runs: Gradual shifts in rainfall polarity**

In this section we test the ability of the second-generation model to extract seasonality information using synthetic data. The model reproduces shifts in rainfall polarity in synthetic datasets well (Figure 7). In one experiment, the input $\delta^{18}O$ dataset was created by using i) a temperature sine function that was set as invariant (i.e., it maintained its polarity and amplitude throughout the run), and ii) a rainfall sine function that shifted in polarity completely over 14 model years. The input sine waves were used to create the annually-resolved synthetic $\delta^{18}O$ record but were independent from the sine waves generated by the model. The wettest month in the input rainfall record was April in Year 1, gradually changing polarity to November by Year 14. As such, model Year 7 was characterised by no seasonality (Figure 7). The model was run without *a priori* knowledge of these shifts other than the mean annually-resolved synthetic $\delta^{18}O$ record, MAT, 'modern' seasonality range, and cave temperature (i.e., the simulations were run 'modeller blind'), but the output reproduced the shifting rainfall pattern very well. The gradual shift in rainfall polarity is detected, and the lack of seasonality in the input rainfall signal during Year 7 is reproduced. The input temperature data had a 15 °C annual temperature range, and two model simulations were conducted: one derived using an annual seasonal temperature range of 10 ± 6 °C, and a second using an annual seasonal temperature range of 15 ± 6 °C. In the case of the lower annual temperature range, the model overestimates rainfall seasonality to compensate for the inappropriate annual temperature range, but still detects shifts in rainfall polarity (Figure 7). When the more appropriate temperature range is used, the simulation captures both the amplitude and polarity of the shifting rainfall input signal. However, this experiment highlights a limitation of this modelling approach; $\delta^{18}O$ data is explicable both in terms of rainfall and temperature seasonality shifts, and an unknown annual temperature range introduces uncertainties.

A second experiment involved synthetic temperature and rainfall input records with both considerable inter-annual variability and noise introduced (Figure 8). Notably, one model year (Year 4) had the polarity of the rainfall signal completely reversed. Again, the model was able to extract the salient features of the input data very well. Reproduced were inter-annual variations in rainfall and temperature, and, importantly, the model detected the reversed seasonality of the rainfall signal in Year 4 (Figure 8).

## 5.2 Application to a stalagmite $\delta^{18}O$ dataset from a seasonally arid continental region

The first version of the model was run successfully across the Holocene using a $\delta^{18}O$ dataset derived from the maritime climate of northern Spain (Baldini et al., 2019). Here, we apply the second-generation model to a dataset from Bir-Uja Cave in the Keklik-Too mountain ridge, Kyrgyzstan, a location characterised by extremely strong seasonal fluctuations in both temperature and rainfall. The cave (40°29'N, 72°35'E) is ~60 m long and is developed at an altitude of ~1,325 m above sea level (Fohlmeister et al., 2017). The input data consisted of the $\delta^{18}O$ dataset from stalagmite Keklik1 reported on in Fohlmeister et al. (2017), a 500-year long, centennial-resolution borehole temperature record from the Tian Shan mountains (~461 km to the north of the cave site) (Huang et al., 2000), instrumental precipitation and temperature records since 1880 C.E. from Tashkent, Uzbekistan (~300 km to the east) (Menne et al., 2012), and cave temperature (Fohlmeister et al., 2017). The $\delta^{18}O$ input data were decadally-resolved, and the stalagmite was dated using a recently developed radiocarbon technique (Fohlmeister and Lechleitner, 2019; Fohlmeister et al., 2017; Lechleitner et al., 2016b). The Keklik1 record extends from 2011 C.E. back to 1150 C.E., but the borehole record only extends back to 1500 C.E., so the interval modelled only extends to 1500 C.E. On average, the site receives ~450 mm of precipitation per year (based on Global Network of Isotopes in Precipitation data from Tashkent), with ~80% falling from November to April. Summers are very dry, with August (the driest month) receiving ~5 mm of rainfall. Monthly temperatures range from -1.4 °C in January to 25.0 °C in July, with a MAT of 12.1 °C. Stalagmite Keklik1 was located ~40 meters from the cave entrance and was collected in October 2011. Cave temperature varies seasonally, from 12 °C from the end of November until April, to a maximum of 16.5 °C in May. The site is characterised by near 100% relative humidity in the cold season which drops considerably to ~60% during the warmer months (Fohlmeister et al., 2017).

Unlike the Spanish GAR-01 record which extended back to ~13,500 years BP and was modelled using 100-year timeslices (Baldini et al., 2019), the Keklik1 $\delta^{18}O$ record was modelled using annual timeslices. The duration of the timeslice is user-defined and is independent of the resolution of the original stalagmite $\delta^{18}O$ dataset, but a timeslice with a somewhat higher resolution than the $\delta^{18}O$ dataset ensures that the input data are entirely represented. The timings of the minimum and maximum values of the modelled temperature sine function were fixed at January and July, respectively. These months were also designated as the minimum/maximum of the modelled rainfall sine wave, which fits present day observations, but the sine function's polarity was not prescribed in advance.

Baldini et al. (2019) noted that the modelled temperature curve for northern Iberia closely resembled a previously published temperature reconstruction for the region (Martin-Chivelet et al., 2011) with a temporal resolution that exceeded the information provided by the low-resolution input dataset. Although no annual-scale MAT record exists in the Kyrgyzstan region for the last 500 years, summer temperatures are well constrained by tree ring records. A comparison of the modelled July temperature derived from the Keklik1 $\delta^{18}O$ record reveals a very good match with the NTREND AG2 temperature anomalies (~300 km to the north of the cave site) (Anchukaitis et al., 2017; Cook et al., 2013) (Figure 9). The model's ability to reconstruct palaeotemperature may reflect the fact that the probability of a successful model run is maximised when modelled temperature approximates the actual temperature shift. Successful model runs with a different (and incorrect) temperature pattern are possible with certain modelled rainfall simulations, but the mean monthly temperature values (reflecting the mean of all successful runs) will be biased towards model simulations with the correct temperature shift. The apparently robust reconstruction of warm-season palaeotemperature is an unexpected and exciting model outcome, but one that requires further evaluation.

The rainfall reconstruction reproduces many of the same features highlighted by Fohlmeister et al. (2017). In particular, decreases in the winter rainfall contributions in the late 1500s, the mid-1700s, and the early 1800s are apparent in both records. This agreement is expected because the $\delta^{18}O$ record is integral to both reconstructions, but it is interesting that the two reconstructions use two fundamentally different techniques (numerical versus geochemical modelling) to estimate the importance of winter rainfall to the overall annual water budget at the site and arrive at broadly similar results. For example, a winter rainfall peak occurs in 1797 CE in both records and transitions to drier winters by 1815 CE, with ~22% and ~50% reductions in winter rainfall implied by the model and $\delta^{18}O$ data, respectively. The model underestimating the reduction in rainfall probably arises because of the model's utilisation of smooth sine waves rather than more step-like functions; in other words, although it is possible for one month per year to have zero rainfall in the model, the adjacent two months must necessarily have some rainfall, whereas in reality, several dry months per summer could occur. The use of step functions would permit the incorporation of several dry months annually and would amplify apparent shifts in seasonal rainfall amounts. Modelled DJFM rainfall compares reasonably well with GHCN rainfall from Tashkent (Figure 9), particularly considering that the Tashkent meteorological station is ~300 km away from and ~1,000 m lower in altitude than the cave site.

### 5.3 Limitations to the modelling technique and future work

Several limitations to the presented modelling technique exist. First, the timing of the rainfall minima and maxima versus temperature signal could affect the model's efficacy; for example, if the rainiest month occurs three months after (or before) the warmest month, the use of the sine function means that all outcomes are possible. This is because the maxima/minima in one parameter's sine function occur at the nodes of the other sine wave, effectively making both sine waves independent of each other. At many sites, temperature and rainfall are intrinsically linked and their seasonal cycle broadly synchronous, but the above may be an issue at some locations. Additionally, the model would require a differently shaped rainfall-function to model rainfall at locations with two distinct rainy intervals every year, such as low latitude sites affected by the ITCZ twice each year.

The current version of the model does not incorporate evapotranspiration, and this is an obvious oversimplification. This may have repercussions for sites like Kyrgyzstan that experience a pronounced hot and dry season with negative effective infiltration. Similarly, variable kinetic fractionation almost certainly occurred within the cave (Fohlmeister et al., 2017) but is not considered within the model. Future versions of the model will incorporate both evapotranspiration and kinetic effects, but the model currently likely overcomes this limitation simply by reducing rainfall amount for months with high evapotranspiration rates. Potentially, coupling the new model discussed here with a dripwater isotope evolution model (e.g., ISOLUTION (Deininger and Scholz, 2019)) could produce very robust results. The model also cannot identify intervals characterized by changes in moisture pathway or fractionation amount; rather, it highlights intervals that are not explicable in terms of changes in temperature or rainfall amount seasonality (intervals where the model cannot converge on any solutions), and thus points to the involvement of other processes.

The model is allowed to randomly vary MAT above or below the low-resolution temperature input record, but only within user-defined bounds. Too great a range of permissible MAT values would allow essentially any outcome. For example, if there were no limits to minimum winter temperature, a low $\delta^{18}O$ value could be modelled as either a very cold winter with a subdued rainfall seasonality or as a mild winter but with substantial winter rain. Limiting the temperature seasonality to reasonable bounds (for example, based modern interannual MAT variability) permits assessing whether any given month is warmer or colder than the low-resolution temperature input, but may underestimate the total amount of cooling and warming. In extreme cases, this may manifest itself as a failure to converge upon any successful model, thus highlighting timeslices that require closer inspection and potentially an alternative explanation.

As discussed in Section 5.2, the utilisation of step functions to describe rainfall seasonality may facilitate the modelling of climate for sites where several months receive similar amounts of rainfall. Future studies should investigate the ramifications of function choice on output. Additionally, theoretically arriving at a mathematical solution utilising the relevant equations and input data is possible, obviating the need for MC simulations, and future research will investigate this possibility. Finally, future models could incorporate options for geochemical modelling of drip and carbonate chemistry.

## 6. Regional seasonality

In this section we analyse global meteoric precipitation and temperature data to highlight regions experiencing pronounced seasonal variability in temperature, precipitation amount, and precipitation $\delta^{18}O$ (Figures 10 and 11), helping to facilitate the identification of cave sites sensitive to seasonality. This also highlights locations that are at the margins of such regions, where seasonality may have affected the record in the past, despite the lack of a modern influence.

### 6.1. Identification of seasonally sensitive regions

WorldClim Version 2 data were obtained at a 2.5 minute (~4.5 km at the equator) spatial resolution (Fick and Hijmans, 2017). Inland continental regions within the mid- to high-latitudes of the Northern Hemisphere (e.g., central and northern Canada, eastern Russia, northeast China, and Mongolia) are characterised by the greatest mean annual temperature range (Figure 10a). A greater annual temperature range is characteristic of continental climates due to the reduced oceanic influence, with ocean water's high heat capacity and moderating influence on air temperature. The lowest mean annual temperature ranges occur in the low latitudes (where insolation remains high year-round) and maritime regions of the world (where oceans moderate temperature variability) (Figure 10a). The pattern of global temperature seasonality (herein calculated as the maximum temperature of the warmest month minus the minimum temperature of the coldest month averaged over the period 1970 – 2000 based on WorldClim Version 2 data) is consistent with the geographic pattern of cave air ventilation reported in (James et al., 2015), a study concerning the role of outside temperature seasonality in the seasonal ventilation of caves.

Seasonality in precipitation amount (Figure 10b) is greatest in the low latitudes due to the annual migration of the Intertropical Convergence Zone (ITCZ) and monsoonal systems that cause distinct wet and dry seasons, along the western coast of North America, southern South America, and Europe where seasonal westerlies preferentially bring enhanced winter precipitation, and bordering the Mediterranean where a 'Mediterranean climate' characterised by wet-winters and dry-summer dominates (Figure 10b). The lowest precipitation amount seasonality occurs in arid and semi-arid regions of the world and the non-coastal mid- to high-latitudes of the northern and southern hemispheres.

Global seasonality in amount-weighted $\delta^{18}O_p$ (Figure 11) approximates the pattern of temperature seasonality (Figure 10a), with the greatest annual range in $\delta^{18}O_p$ observed at Northern Hemisphere continental interior and high latitude sites (e.g., northeast Asia, central Canada, northern Greenland). In addition, high altitude sites (e.g., the Andes in western South America, the Caucasus Mountains at the intersection of Europe and Asia) also exhibit higher annual WM $\delta^{18}O_p$ ranges due to the altitude effect. The lowest $\delta^{18}O_p$ seasonality occurs within maritime (e.g., NW Europe, SW and SE Australia) and arid/semi-arid regions (e.g., East Africa, eastern Brazil, South Africa). Many stalagmite records are from temperate regions where modern MAT ranges from 10 to 16 °C (Baldini et al., 2019; Baldini et al., 2015; Ban et al., 2018; Huang et al., 2001; Johnson et al., 2006; Orland et al., 2014). Global cave dripwater $\delta^{18}O$ data reveal that caves from regions with this MAT range have dripwater chemistry that reflects recharge-weighted $\delta^{18}O_p$ (Baker et al., 2019). The seasonal distribution of $\delta^{18}O_p$ is therefore a critical control in the case of many different stalagmite samples.

In other cases, very pronounced seasonality inherent in stalagmite geochemical records are not due to seasonality in $\delta^{18}O_p$, but instead to seasonality in rainfall amount (Ridley et al., 2015b) and associated shifts in bioproductivity (Baldini et al., 2005) or PCP (Fairchild and Hartland, 2010; Fairchild et al., 2006). Seasonality in temperature can also induce cave ventilation in temperate zone caves during the winter (providing the cave geometry is appropriate), promoting carbonate deposition within the cave and biasing annual- to decadal-scale records towards the winter season rainfall (James et al., 2015). The maps provided herein can help identify regions containing speleothems retaining the desired seasonal signal, and determine what the most likely control is on any seasonal signal found within a stalagmite. Furthermore, the maps help highlight cave sites that are located on the peripheries of climatologically seasonal zones at present, where past seasonality shifts could have influenced a record. Examples include the Sahel and southern Belize (Figure 12), both currently at the very northern extent of the ITCZ, where a small ITCZ shift to the south would produce both severe drying and a substantial decrease in rainfall seasonality. This perspective was underscored by recent results from Central America that used monthly-scale rainfall proxy data over the last two millennia to suggest that the region has only been affected by the ITCZ since ~1400 C.E., and that the ITCZ influence may wane in the near future (Asmerom et al., 2020) (Figure 12).

**6.2. Complexities despite strong seasonality: northeast India as an example**

The seasonality maps presented here highlight regions most likely to contain stalagmites which retain seasonal signals in temperature, rainfall amount, or $\delta^{18}O_p$. However, they also illustrate that not all seasonal variations in $\delta^{18}O_p$ are explicable in regional temperature or rainfall amount terms. In many cases, complex moisture source variability overprints temperature-induced seasonality, hampering the use of models such as the one presented in Section 5. Here, we discuss the Indian Summer Monsoon (ISM) as an example of such a situation, and focus specifically on Mawmluh Cave in Meghalaya, northeast India, one of the most seasonal locations on Earth in terms of rainfall amount (Fig. 10). In Meghalaya, hydroclimate is characterised by extreme seasonality, as the plateau constitutes the first topographic barrier for moisture-laden air masses travelling inland from the Bay of Bengal (Murata et al., 2007; Prokop and Walanus, 2003). At present, the ISM brings ~80% of the annual rainfall to the cave site, inducing extreme amounts of rainfall (up to 12 meters per year (Breitenbach et al., 2015). The seasonal precipitation cycle is reflected in rainfall $\delta^{18}O$ composition (Berkelhammer et al., 2012; Breitenbach et al., 2010). Rainfall $\delta^{18}O$ becomes progressively lighter during the ISM, but this effect is only partially driven by increasing precipitation intensity and the amount effect because the period of maximum precipitation (June-August) precedes maximum $^{18}O$ depletion (August-October) (Breitenbach et al., 2010)). Instead, the $^{18}O$-depletion results predominantly from the moisture source shifting from a proximal location (the Bay of Bengal) in the early and late ISM to a more distal location (the open Indian Ocean) during the peak ISM (longer transport times resulting in more Rayleigh distillation). Rainfall and dripwater $\delta^{18}O$ at Mawmluh Cave are thus highly seasonal, but the relationship between temperature, rainfall amount, and rainfall $\delta^{18}O$ is not straightforward (Breitenbach et al., 2010; Breitenbach et al., 2015). Additional complexity arises from the filtering and buffering capacity of the karst aquifer through which rainwater percolates *en* *route* to a stalagmite. Although a clear seasonal dripwater $\delta^{18}O$ cycle exists, with its lowest value approximating ISM rainfall $\delta^{18}O$, its annual amplitude is compressed, reflecting buffering in the karst (Breitenbach et al., 2015). This further complicates the interpretation of $\delta^{18}O$ records from these stalagmites, and information from independent proxies that are sensitive to processes dominating during the winter season is required to disentangle such processes. Combining summer-sensitive $\delta^{18}O$ with winter-sensitive Mg/Ca (reflecting PCP) permitted disentangling ISM strength and the degree of dry season dryness in a stalagmite from Mawmluh Cave (Myers et al., 2015; Ronay et al., 2019). Such a multi-proxy approach, supported by local monitoring and karst process modelling, allows robust interpretations of seasonal-scale climate from stalagmites, even when the proxy seasonality is driven by more complex processes than temperature or rainfall amount alone.

## 7. Future directions and recommendations

In this review, we introduce and discuss several concepts that we hope will facilitate the development and interpretation of robust seasonal-resolution climate records from stalagmites, will improve the extraction and interpretation of seasonal information from stalagmites, and promote future discussion, including: **A)** that replication of records should not always be an expectation without *a priori* knowledge that the drip type and environmental conditions responsible for the deposition of the stalagmites are comparable (e.g., some stalagmites retain seasonal information, others do not), **B)** that every stalagmite-based geochemical record is different and records a unique component of the environmental signal of varying complexity (i.e., each stalagmite retains an accurate history of its environment; the question is whether or not this history can be deconvolved), and **C)** that the application of at least one year's worth of hourly-resolved drip rate monitoring combined with a new drip classification scheme presented here may help identify stalagmites retaining a seasonal signal. Furthermore, we have **(D)** developed global seasonality maps of temperature (as was done previously by (James et al., 2015)), meteoric precipitation amount, and meteoric precipitation $\delta^{18}O$ ratios which allow the identification of regions sensitive to different types of seasonality recordable by stalagmites. The maps facilitate predicting what type of seasonality potentially affects modern stalagmite samples from that region. They also assist in palaeoclimate interpretations by identifying locations proximal to regions with pronounced seasonality, where past migration of key atmospheric circulation systems could have altered the geochemical record retained by a stalagmite. On a similar note, we **(E)** present a model that interprets annual- to centennial-scale stalagmite $\delta^{18}O$ records in terms of seasonal temperature and meteoric precipitation seasonality shifts. Although we stress that this model only highlights one possible interpretation (that the data were modulated primarily by regional long-term mean annual temperature variability combined with seasonality shifts in rainfall and temperature), often this interpretation is the most parsimonious. The modelling technique also helps identify time intervals when altered seasonality cannot account for the observed isotope shifts, suggesting that another variable needs consideration. We **(F)** discuss four major controls on the seasonality signal within stalagmites: i) Earth atmospheric, ii) Meteoric precipitation, iii) biological (e.g., soil processes), and iv) cave atmospheric, and **(G)** discuss a case study from India that serves as an example of a stalagmite whose seasonal signal is not derived from rainfall amount or regional temperature, but instead results from seasonal shifts in air mass trajectories (i.e., affected by seasonal shifts in Earth atmospheric processes).

Stalagmites are remarkable archives of information regarding climate (on both seasonal and longer timescales), surface and cave environmental conditions, dry deposition, moisture source pathway, marine aerosols contributions, and hydrological routing. Replication of proxy records present strong support for palaeoclimatic interpretations and should remain a goal of any stalagmite science research programme, but unless the climate signal-to-noise ratio of a region is unusually high, replication is only possible when comparing stalagmites deposited under similar conditions. A thorough understanding of the environmental processes affecting both entire caves (e.g., ventilation) as well as individual stalagmites (e.g., drip rate) facilitates replication efforts. The geochemical record from even adjacent stalagmites will reflect numerous processes, some of which are common to the two samples but many which are not, and only through a thorough understanding of the processes affecting each sample are robust (and replicable) climate interpretations achievable. However, unless analytical issues exist, non-replication does not imply that one record is incorrect; rather it generally implies that the two records simply record different environmental parameters.

Cave monitoring prior to the collection of a stalagmite will increase the likelihood of obtaining a record of the desired sensitivity to seasonal climate shifts, or other desired forcing. We recommend monitoring the drip feeding the stalagmite for at least one year using an automated drip logger and plotting the results in a diagram similar to Figure 3 to evaluate a stalagmite's likelihood of retaining hydrological seasonality. We recommend monitoring multiple sites within the cave and selecting the most appropriate stalagmite for collection based on the monitoring results. It is worth bearing in mind that unless the seasonality signal in a stalagmite is conveyed via seasonal cave ventilation, stalagmites fed by diffuse flow drips with long residence times may not retain seasonal information. Other drips that are seasonally either dry or undersaturated with respect to carbonate will lead to the occurrence of seasonal hiatuses in the stalagmites and signal loss for that particular season. Monitoring a stalagmite's drip rate and drip chemistry for as long as possible represents one of the simplest but most effective means of understanding the potential climate signal contained within a sample prior to collection. This also has implications for cave conservation and protection efforts, because clearly formulated research goals and drip monitoring prior to stalagmite sample collection can greatly reduce the number of samples removed from a cave for research purposes.

If sample growth rate permits, we suggest that the extraction of the palaeoseasonality signal over millennial timescales is best achieved via micromilling, leaving no gap between adjacent samples, or LA-ICPMS. The major disadvantages of micromilling are that it is resource intensive and that many samples may not have growth rates high enough to permit the required temporal resolution. The major disadvantage of LA-ICPMS is that the trace element signature of a stalagmite is often dominated by site-specific factors such as temperature, sea spray, volcanic aerosols, fire, variable throughput of colloidal material, or rainfall, and consequently aligning the data with other records is sometimes complex. Micromilled carbonate powders that are divided into two or more aliquots that are subsequently analysed for stable isotope ratios, trace elements, and other geochemical proxies can provide very robust interpretations (e.g., Jamieson et al., 2016). This eliminates issues of cross-correlation and enables a powerful multiproxy approach, where each stable isotope ratio value is linked directly and unambiguously to numerous elemental concentration values. The technique can yield important information regarding palaeoseasonality but is considerably more resource intensive than running multiple LA-ICPMS tracks parallel to each other and the micromilled stable isotope track. An alternative is to produce a long decadal-scale isotope ratio traverse complemented by higher resolution transects or maps across key intervals of interest using LA-ICPMS, SIMS, synchrotron, or µXRF to corroborate interpretations based on the longer transects. In the future, proxy mapping at micron-scale resolution using these techniques will help reduce uncertainties related to geometric ambiguities such as those associated with crystal boundaries and improve the robustness of interpretations.

## 9. Conclusions

The reconstruction of palaeoseasonality using stalagmites is an exciting research direction that has yet to mature into its full potential. Numerous records of palaeoseasonality exist, but few direct reconstructions extend before the last two millennia. Ideally, future studies concluding that a decadal- to annual-scale isotope ratio record is affected by seasonality changes should support this by either using short windows of sub-annual data or by modelling.

Any stalagmite-based climate proxy record is affected by inherent complexities in climate signal transfer to the stalagmite and by selective sampling of the stalagmite for analysis. A high-resolution (sub-annual to annual-scale) sampling strategy coupled with appropriate site monitoring maximises the likelihood of extracting a signal approximating the climate input signal. For long records annual- to decadal-scale resolution is ideal, and shorter records could benefit from an even higher resolution if resources permit. Large shifts in isotope ratios could reflect changes in seasonality, potentially associated with the migration of key atmospheric circulation systems over the cave site. New models incorporating seasonality can provide information regarding whether observed geochemical shifts are interpretable in terms of altered seasonality, and these represent an exciting and inexpensive new research tool. A seasonal-scale sampling strategy over short intervals of interest can verify these model interpretations, and LA-ICPMS or line-scan µXRF represent potentially the most efficient methods to achieve this; other alternatives include monthly-scale micromilling, synchrotron analysis (SR-µXRF), and SIMS.

The robust interpretation of stalagmite geochemical records in terms of seasonality represents a key challenge for the next decade. Achieving this is complicated by multiple in-cave and exogenic environmental forcings with dynamic seasonality, including: rainfall, temperature, humidity, bioproductivity, cave air $p$CO$_2$, drip rate, source moisture region and $\delta^{18}$O, and moisture mass trajectory from the source region. Even apparently straightforward $\delta^{18}$O records from regions with high signal-to-noise ratios typically interpretable as either varying total annual rainfall or summer rainfall may reflect another parameter instead (e.g., a change in moisture source or rainfall seasonality), as is the case with the Indian Summer Monsoon. Most records would benefit from a rigorous multi-proxy approach utilising not only multiple geochemical proxy datasets, but also site monitoring and new modelling approaches. Similarly, focussing research efforts at the same well-understood cave sites both maximises the quality of interpretations and contributes to the conservation of caves and stalagmite samples. The application of multiple stalagmites from the same site but with different drip rates and affected by different amounts of disequilibrium fractionation may provide the key to reconstructing formerly elusive climate variables, such as temperature. Instead of representing an irresolvable issue, we suggest that disequilibrium fractionation may present opportunities to quantify temperature, potentially even at seasonal resolutions. Similarly, multi-proxy data could yield seasonal information even in the absence of seasonal sampling resolution; if two or more independent proxies reflect different seasonal data, combining the proxies could yield palaeoseasonality.

Over the past few decades stalagmites have provided some of the most iconic records in palaeoclimatology. In the future, stalagmites will continue to not only provide long records of exceptional quality, but they will also provide rare glimpses into palaeoseasonality at unprecedented temporal resolution. Recent microanalytical advances have facilitated the construction of exquisitely resolved stalagmite-based climate records; we are now at a stage where the interpretation of these records is catching up with their remarkable technical aspects. Extracting quantitative and accurate seasonal climate information from these geochemical records is a key challenge over the next decade, and, if this is achieved, stalagmites will truly be considered in a class of their own as climate archives.

**Acknowledgements**

We thank SISAL and PAGES for access to the SISAL database v1b. Portions of this research were funded by European Research Council Grant #240167. Tim Horscroft is thanked for his support in facilitating the preparation of the manuscript. Ian Orland and Jasper Wassenburg are thanked for detailed constructive reviews that greatly improved the manuscript. Alex Iveson is thanked for useful comments regarding LA-ICPMS.

Iberian temperature and rainfall seasonality over the Younger Dryas and Holocene. Quaternary Sci. Rev. 226, 105998.

Baldini, L.M., McDermott, F., Baldini, J.U.L., Arias, P., Cueto, M., Fairchild, I.J., Hoffmann, D.L., Mattey, D.P., Müller, W., Nita, D.C., Ontañón, R., Garciá-Moncó, C., Richards, D.A., 2015. Regional temperature, atmospheric circulation, and sea-ice variability within the Younger Dryas Event constrained using a speleothem from northern Iberia. Earth Planet. Sci. Lett. 419, 101-110.

Ban, F.M., Baker, A., Marjo, C.E., Duan, W.H., Li, X.L., Han, J.X., Coleborn, K., Akter, R., Tan, M., Nagra, G., 2018. An optimized chronology for a stalagmite using seasonal trace element cycles from Shihua Cave, Beijing, North China. Scientific Reports 8, 4551.

Banner, J.L., Guilfoyle, A., James, E.W., Stern, L.A., Musgrove, M., 2007. Seasonal variations in modern speleothem calcite growth in Central Texas, USA. J Sediment Res 77, 615-622.

Bar-Matthews, M., Ayalon, A., Matthews, A., Sass, E., Halicz, L., 1996. Carbon and oxygen isotope study of the active water-carbonate system in a karstic Mediterranean cave: Implications for paleoclimate research in semiarid regions. Geochim. Cosmochim. Acta 60, 337-347.

Bergel, S.J., Carlson, P.E., Larson, T.E., Wood, C.T., Johnson, K.R., Banner, J., Breecker, D.O., 2017. Constraining the subsoil carbon source to cave-air $CO_2$ and speleothem calcite in central Texas.  217, 112-127.

Berkelhammer, M., Sinha, A., Stott, L., Cheng, H., Pausata, F.S.R., Yoshimura, K., 2012. An abrupt shift in the Indian Monsoon 4000 years ago. Geophysical Monograph Series 198, 75-87.

Blyth, A.J., Baker, A., Thomas, L.E., Van Calsteren, P., 2011. A 2000-year lipid biomarker record preserved in a stalagmite from north-west Scotland. J. of Quaternary Sci. 26, 326-334.

**Figure Captions:**

Figure 1: Top Panel: Resolution of speleothem isotope records over time, compiled from the SISALv1b database. Individual record resolution (small black circles) and mean resolution of all available (black bars) and Holocene (blue bars) records published in a given year. Bottom panel: Total number of stalagmite records identified (grey bars), total number of stalagmite records in SISALv1b (black bars), and total number of Holocene records in SISALv1b (blue bars).

Figure 2: Illustration of different drip responses from Yok Balum Cave, Belize, over approximately two months as captured by a series of automated drip loggers. Two clear rain events and the subsequent drip responses are indicated by the vertical dashed red lines. Rainfall amount is recorded directly over the cave site using a tipping bucket rain gauge. Techniques are discussed in more detail in (Ridley et al., 2015a).

Figure 3: A new drip categorisation scheme designed to emphasise cave drip seasonality. The scheme does not use classification boundaries as such, but instead uses the data distribution to understand the hydrology. The scheme uses descriptors that map onto established drip terminology (see Panels B-D and main text for examples). A) Minimum and maximum hourly drip rates extracted for every month of record for numerous cave drips globally. The dashed line represents the 1:1 line, and all data points must necessarily plot over this (i.e., the minimum drip rate cannot exceed the maximum drip rate for any given month). The closer a point plots to the dashed line, the lower the difference between monthly maximum and minimum values for that point; if a point sits on the line the minimum and maximum values for that month are identical. Panels B-D illustrate some common drip types (using synthetic data) and their pattern when plotted on this diagram. Panels B-D are schematic and are not based on actual collected datasets; the symbols used are arbitrary and are not linked to the symbols used in Panel A.

Figure 4: The simulated effects of sampling resolution on the climate signal extracted from a stalagmite. The stalagmite data are from stalagmite YOK-G (Yok Balum Cave, Belize), which was originally sampled with a micromill at a 100 micron (0.1 mm) step size (Ridley et al., 2015b). The chronology for the stalagmite is precise at the seasonal scale. The rainfall data

(bottom panel) are from the Punta Gorda meteorological station (~30 km to the southeast of the cave site).

Figure 5: Schematic of a sampling scheme for achieving ~50 micron spatial resolution. Plan view of a stalagmite surface with 1 mm conventional holes on the right and trenches cut for low and high resolution. The red trench was milled with a 0.8 mm diameter drill and the (blue-shaded) higher resolution trench was cut laterally, with each sample integrating 50 µm. The red corners highlight the area that is incorporated into subsequent steps, which in this case includes material from the current and the previous sample. In this example each high-resolution sample (e.g., yellow shaded area) integrates a minimal amount of powder of an older sample (because the milling direction is upward).

Figure 6: Several examples of output generated by different geochemical-based techniques for extracting seasonal climate. A) Variability in sulphate in speleothem calcite (Obi84, Obir cave, Austria) as determined by SR-µXRF (Wynn et al., 2014). The clear annual sulphur maxima are evident as brighter green colours. B) Ion microprobe-resolved strontium and phosphorous cycles apparent in stalagmite CC3 from Crag Cave, southwestern Ireland (Baldini et al., 2002). The well-developed cycles illustrate stronger seasonality at the time of deposition (~8.336 ka BP) than currently present. C) Annual UV-luminescent banding in a stalagmite from Shihua Cave, Beijing, China (adapted from Tan et al. (2006)). D) well-develop carbon isotope ratio cycles in stalagmite YOK-G from Yok Balum Cave, Belize, constructed using data obtained via micromilling at a 100-micron spatial resolution and analyses of powders on an IRMS (Ridley et al., 2015b) (see also Figure 4). E) Mg cycles apparent in stalagmite BER-SWI-13 from Leamington Cave, Bermuda, resolved using LA-ICPMS-derived Mg data (Walczak, 2016). All panels show three to four cycles, interpreted as annual.

Figure 7: A synthetic rainfall input signal (orange circles) with an annual temperature range of 15 °C compared with two mean model outputs, one derived using an annual temperature range of 10 ± 6 °C (grey line), and another derived using an annual temperature range of 15 ± 6 °C (blue line). At the beginning of the simulated rainfall input signal record (year = 0), April is the wettest month and November the driest month, but this shifts in polarity slowly through the record, moving through a brief phase with no seasonality in rainfall (year = 7), and then transitioning into a phase where April is the driest month (from year = 8). The vertical gridlines highlight the month of April during every model year. The simulated rainfall input signal amplitude and polarity is reproduced by the model very satisfactorily, provided that the model temperature range is realistic, as it is in Model 2. Note that the polarity of the simulated rainfall input signal is still reproduced by Model 1, but modelled rainfall seasonal amplitude is too large in order to compensate for the low amplitude of the modelled temperature range.

Figure 8: Temperature (top panel) and rainfall (bottom panel) modelling results (black dashed lines) against 'noisy' synthetic input datasets (solid coloured lines) for seven model years. The grey rectangle highlights one model year (Year 4) where the input rainfall signal polarity was reversed; the model detects this shift. The modelling results presented are the mean values of all successful model runs for each timeslice.

Figure 9: Mean modelled monthly temperature and rainfall data against Global Historical Climate Network (GHCN) and tree ring data. A) Stalagmite Keklik1 oxygen isotope ratio data from Bir-Uja Cave, Kyrgyzstan (input data) (Fohlmeister et al., 2017). B) Centennial-scale borehole temperature data from the Tian Shan region (Huang et al., 2000) from 1500 to 2000 C.E. (input data, shifted upwards for clarity) (blue diamonds), modelled July temperature (black curve) (output), and NTREND summer temperature reconstruction for Asia Grid 2 (AG2) (red curve) (Cook et al., 2013). C) Modelled January rainfall (black curve) (output) and GHCN January rainfall for Tashkent (orange curve), both in % of total annual rainfall. The grey rectangles highlight the years 1797 and 1815 C.E. discussed in the text.

Figure 10: Global seasonality in annual temperature (°C) and annual precipitation (mm). A) The annual temperature range was calculated as the maximum temperature of the warmest month minus the minimum temperature of the coldest month averaged over the period 1970-2000. B) Precipitation seasonality was calculated as the precipitation amount of the wettest month minus the precipitation amount of the driest month averaged over the period 1970-2000. WorldClim Version 2 data (https://www.worldclim.org/) were obtained at a 2.5 minute (~4.5 km at the equator) spatial resolution (Fick and Hijmans, 2017). The data span the period 1970-2000 and thus may reflect anthropogenically-influenced temperature seasonality as discussed in Santer et al. (2018). Therefore, although the general spatial pattern of temperature (and potentially precipitation) seasonality may persist into the past, the magnitude of seasonality shifts may deviate from that presented here, particularly when extending records into the preindustrial era.

Figure 11: Global seasonality in amount-weighted precipitation $\delta^{18}O$ (‰ VWMOW). The amount-weighted mean (WM) monthly precipitation $\delta^{18}O$ data (IAEA/WMO, 2001) were used to determine the annual range in precipitation isotopes globally (calculated as the maximum monthly WM $\delta^{18}O$ minus minimum monthly WM $\delta^{18}O$ at 267 stations (yellow symbols) with a complete 12-month dataset over the period 1961-1999. GNIP station data were interpolated onto a 2.5° X 2.5° global grid (~278 km X 278 km) (IAEA, 2001).

Figure 12: A Hovmöller plot of the annual cycle of total-column precipitable water vapour for Central America, based on daily ERA5 re-analysis data across the region from -110 to -

80W and 0 to 35N for the period 1979-2018. Also indicated are the latitudes of three key cave sites that have yielded stalagmites which have produced oxygen isotope records of rainfall.

Figures:

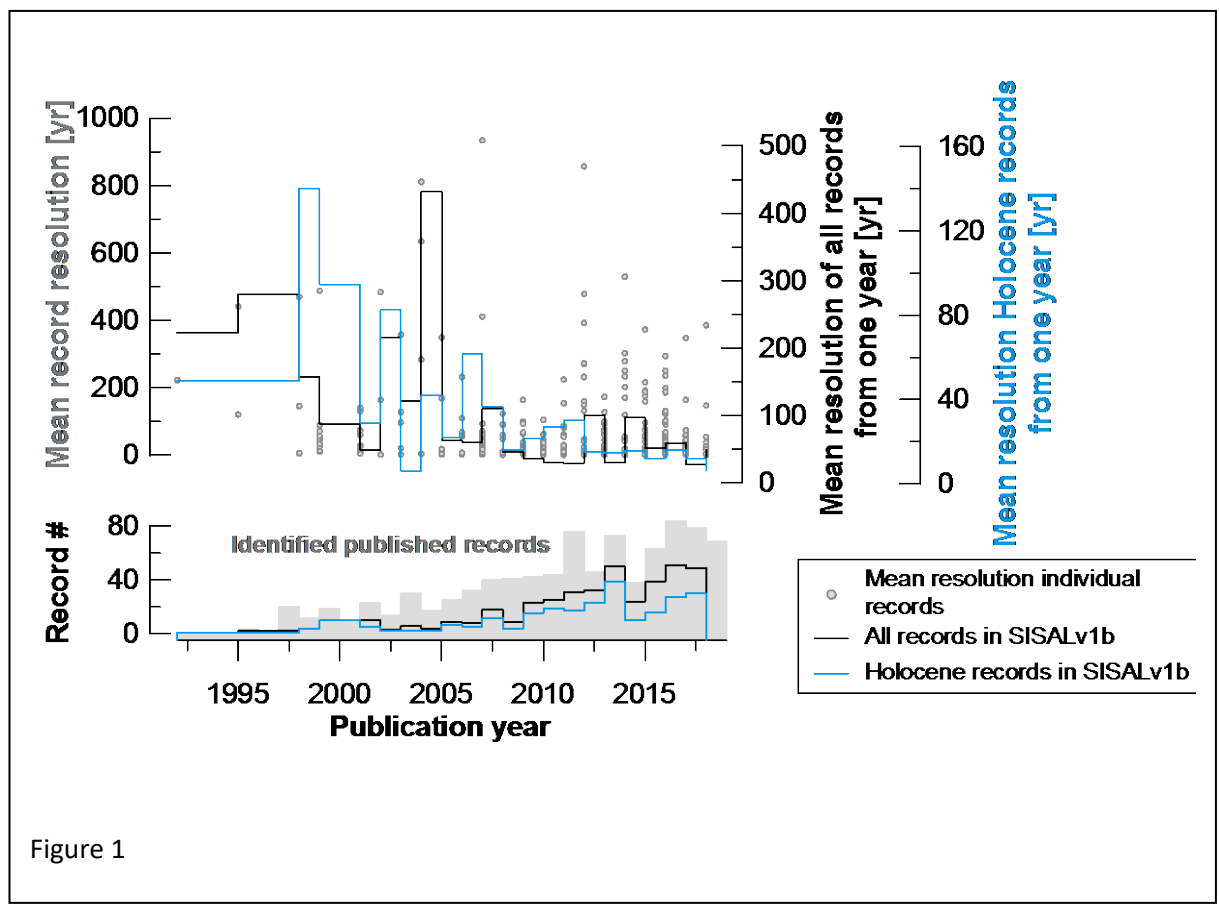

Figure 1

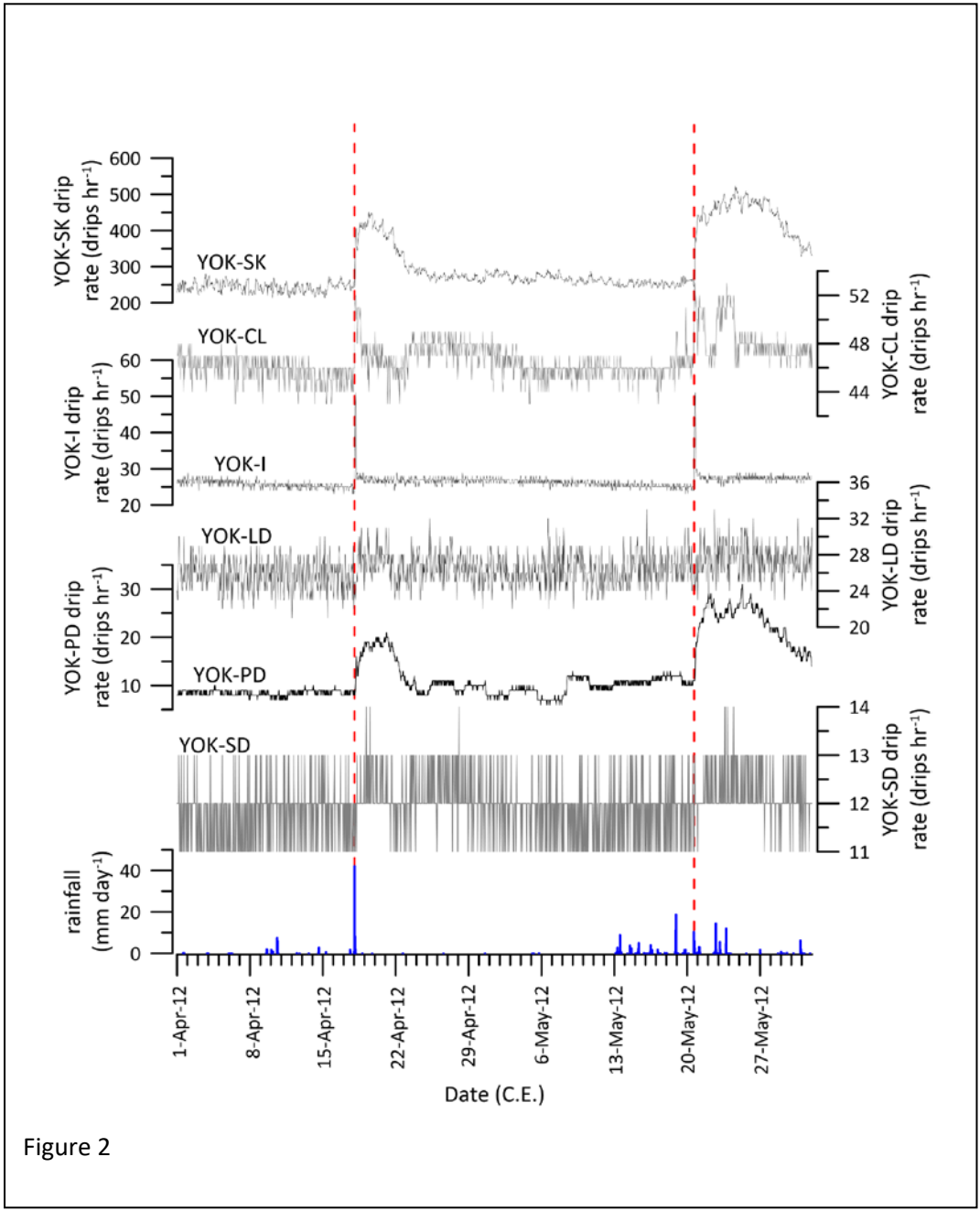

Figure 2

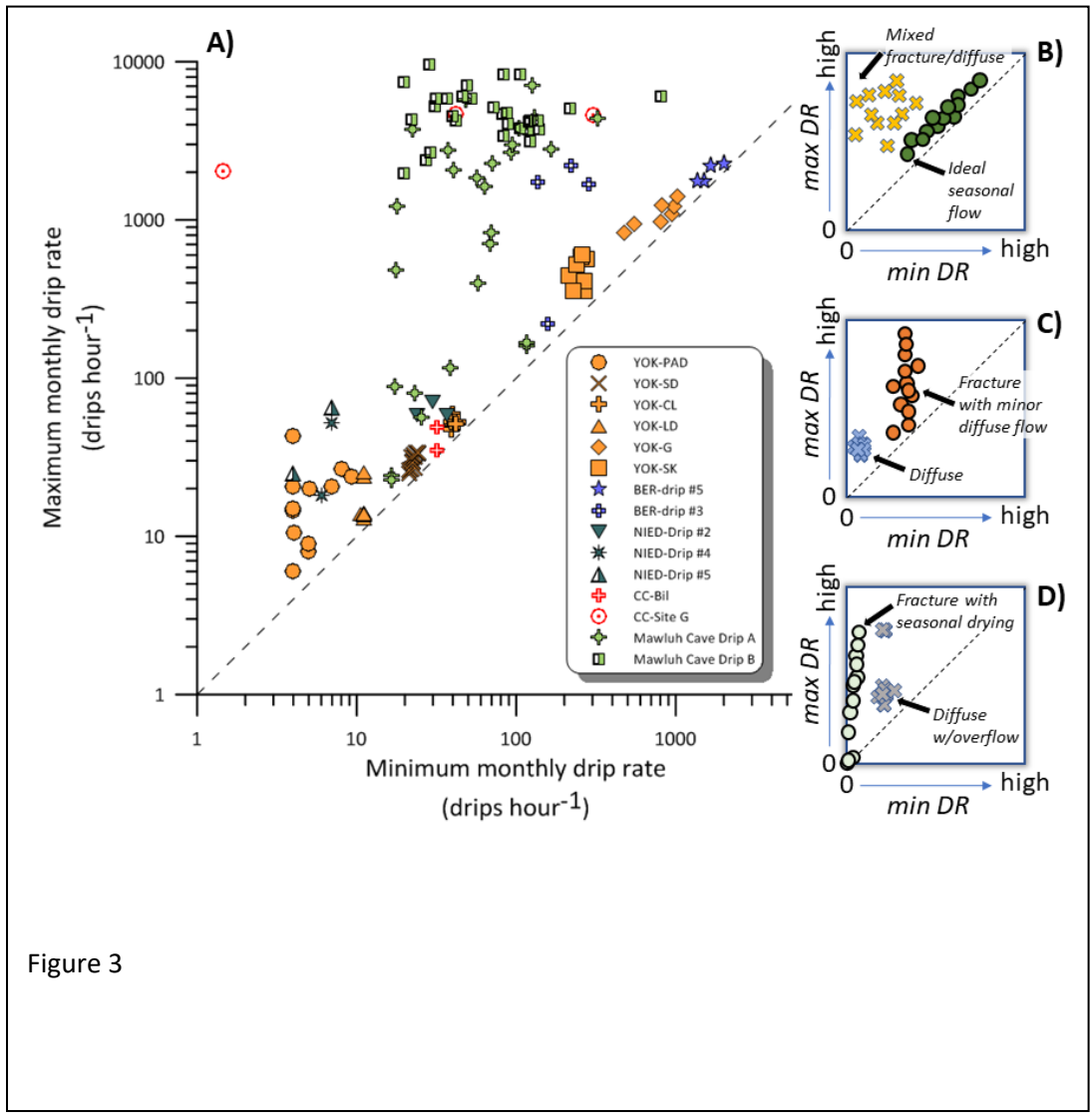

Figure 3

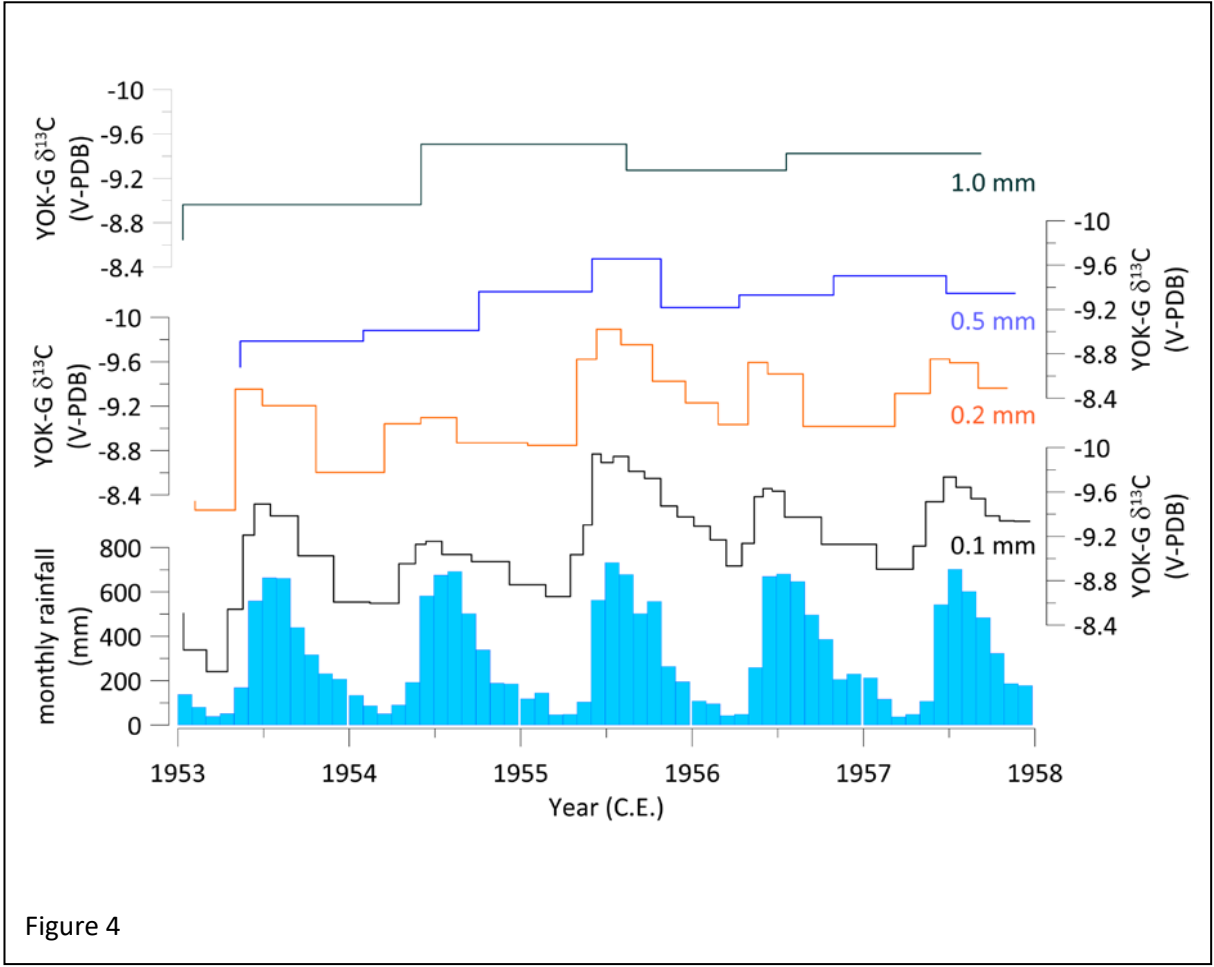

Figure 4

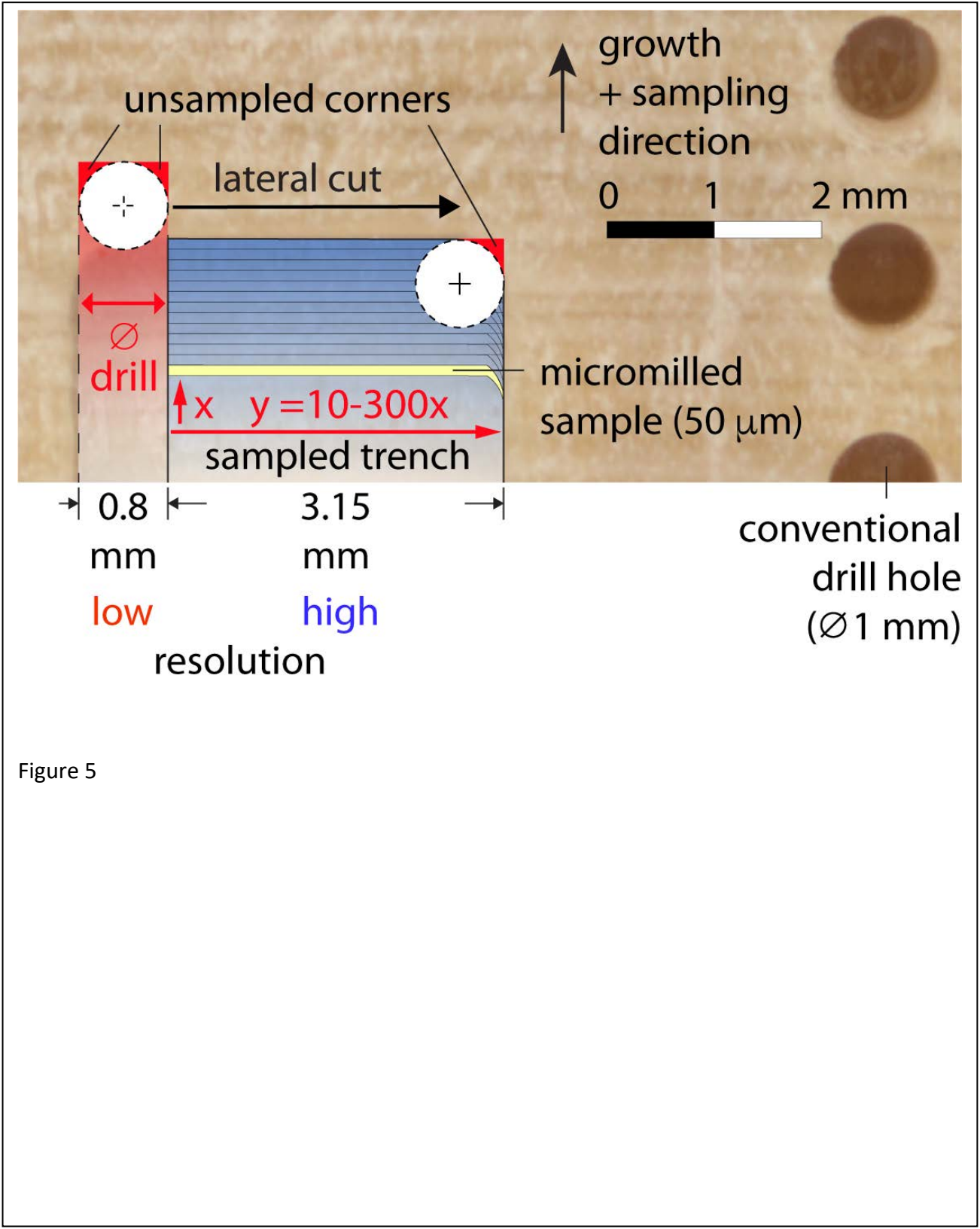

Figure 5

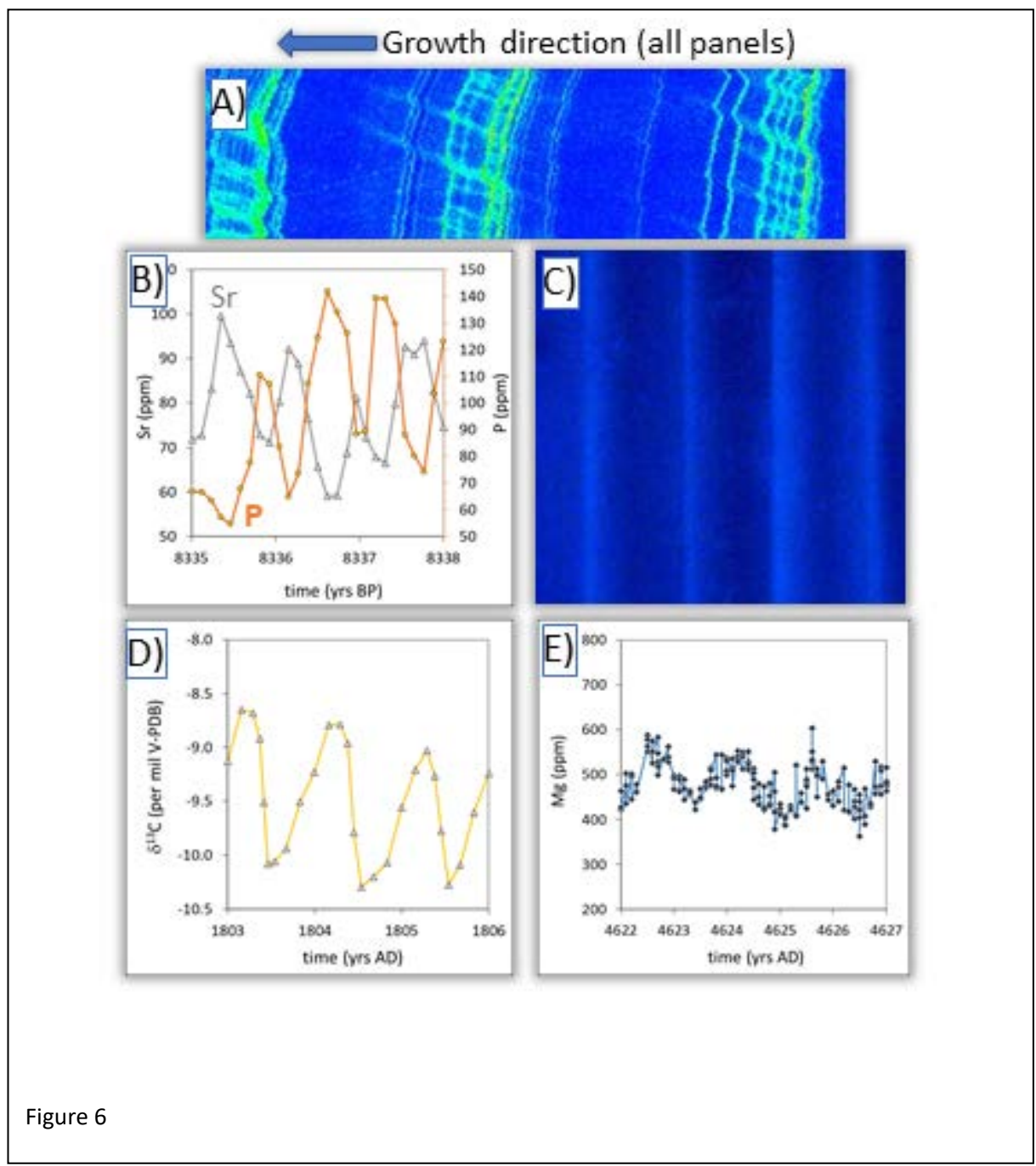

Figure 6

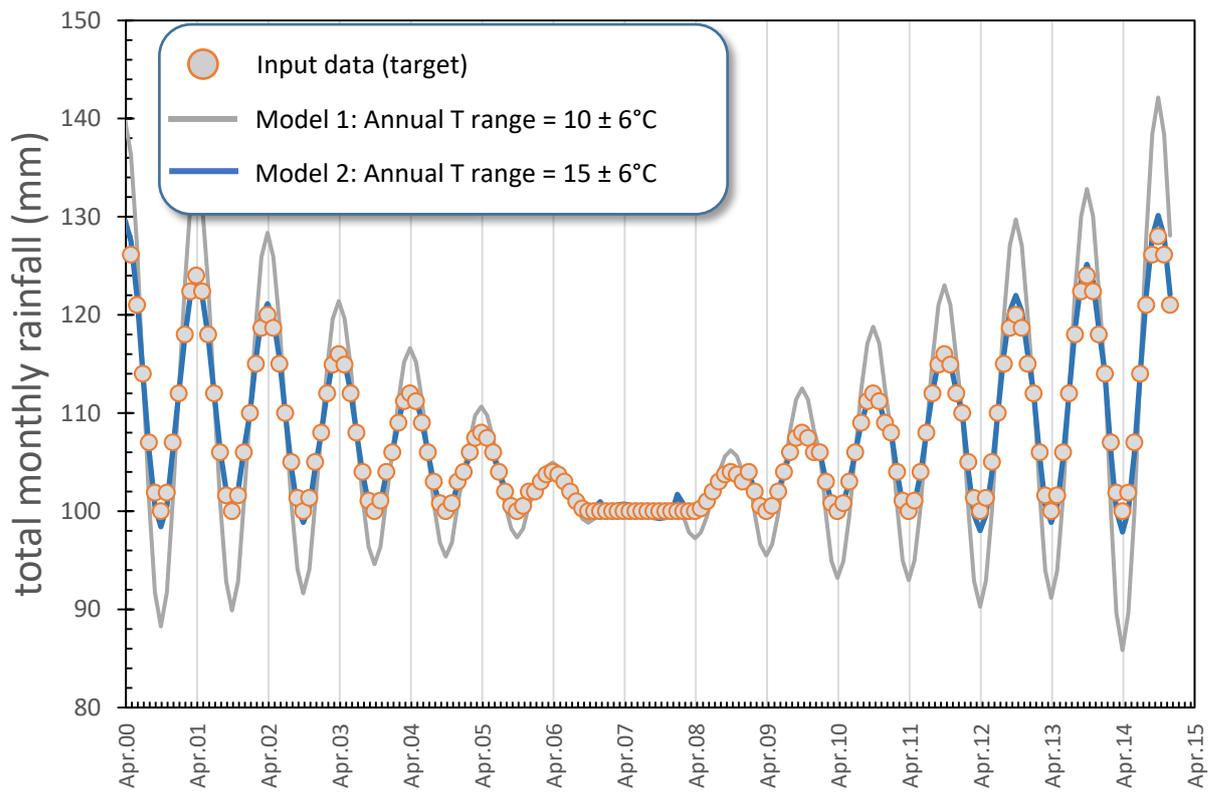

Figure 7

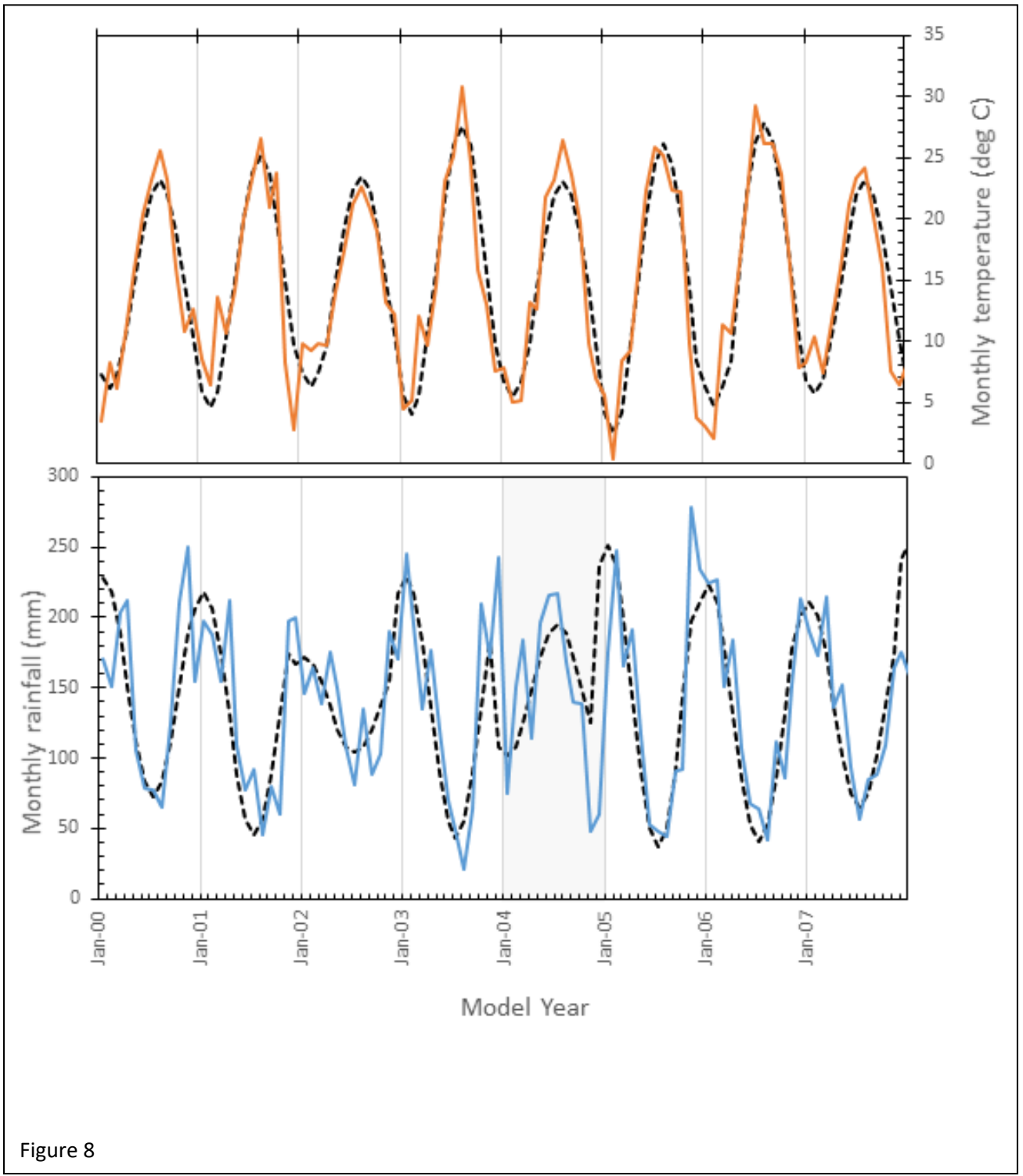

Figure 8

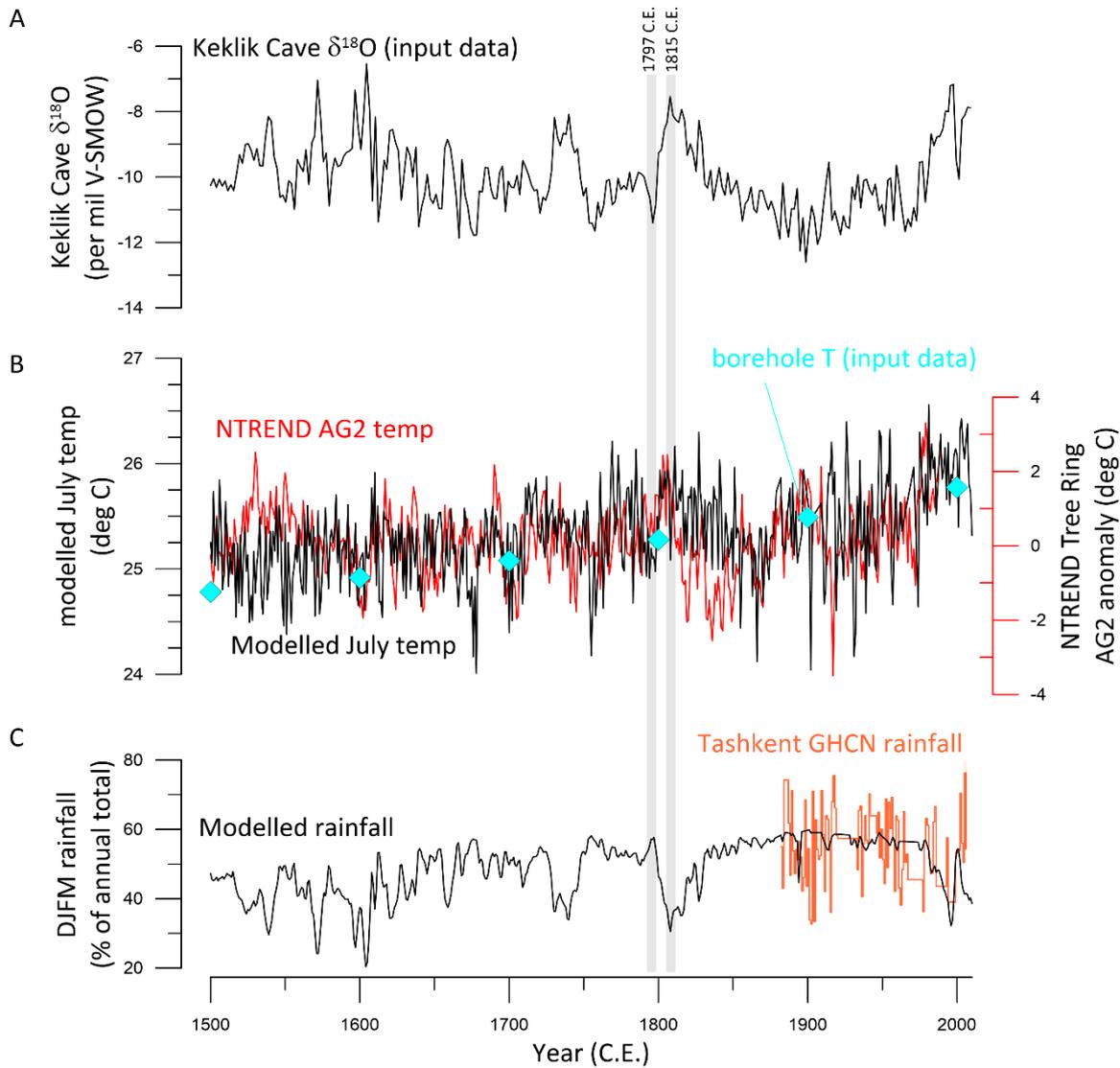

Figure 9

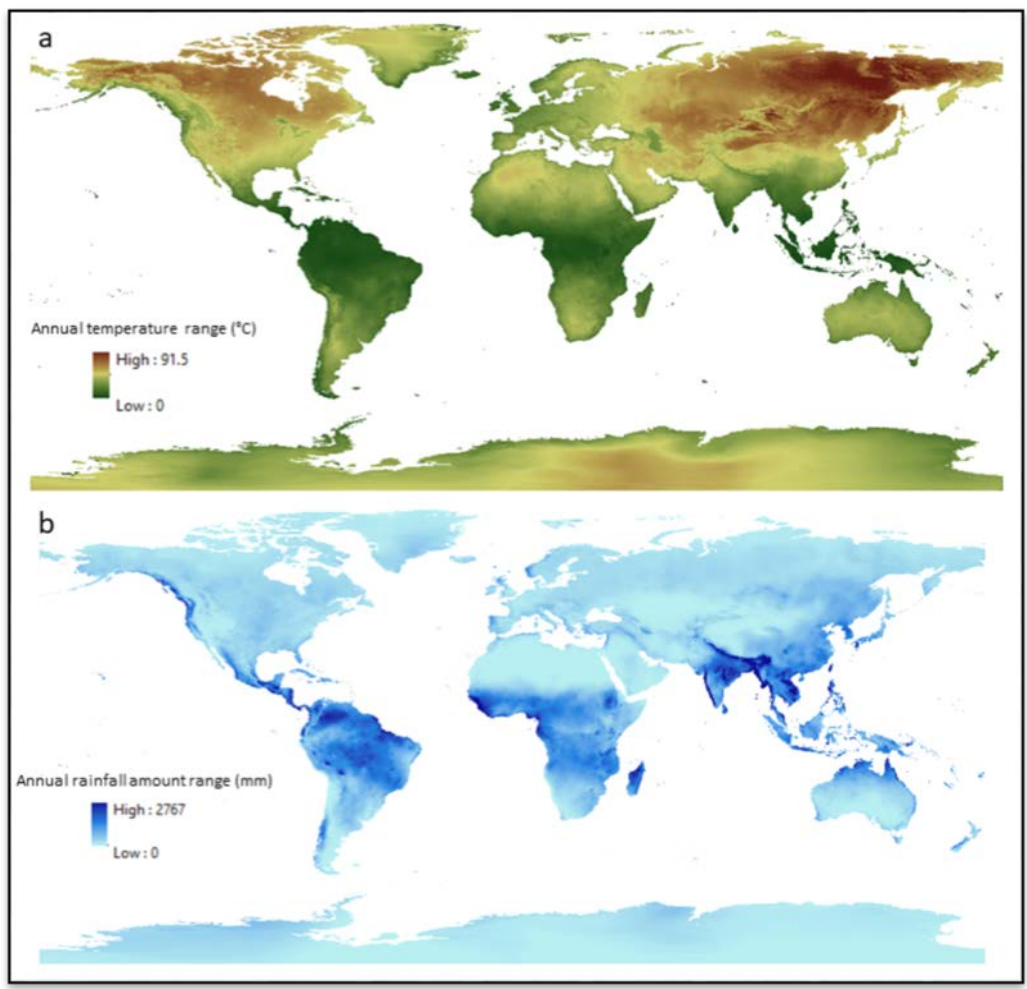

Figure 10

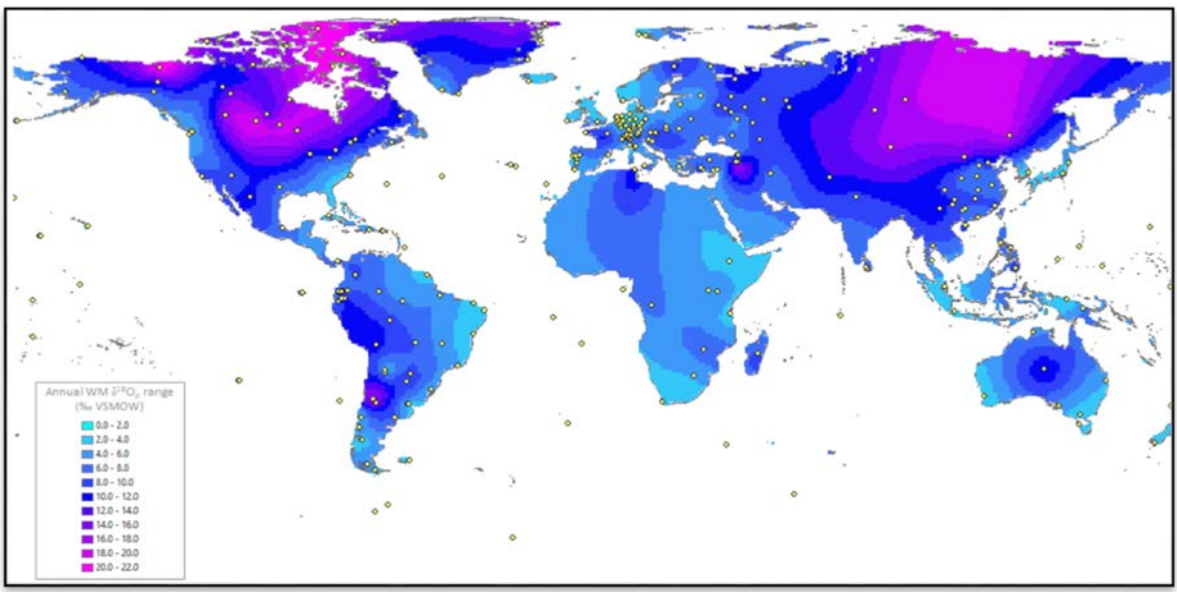

Figure 11

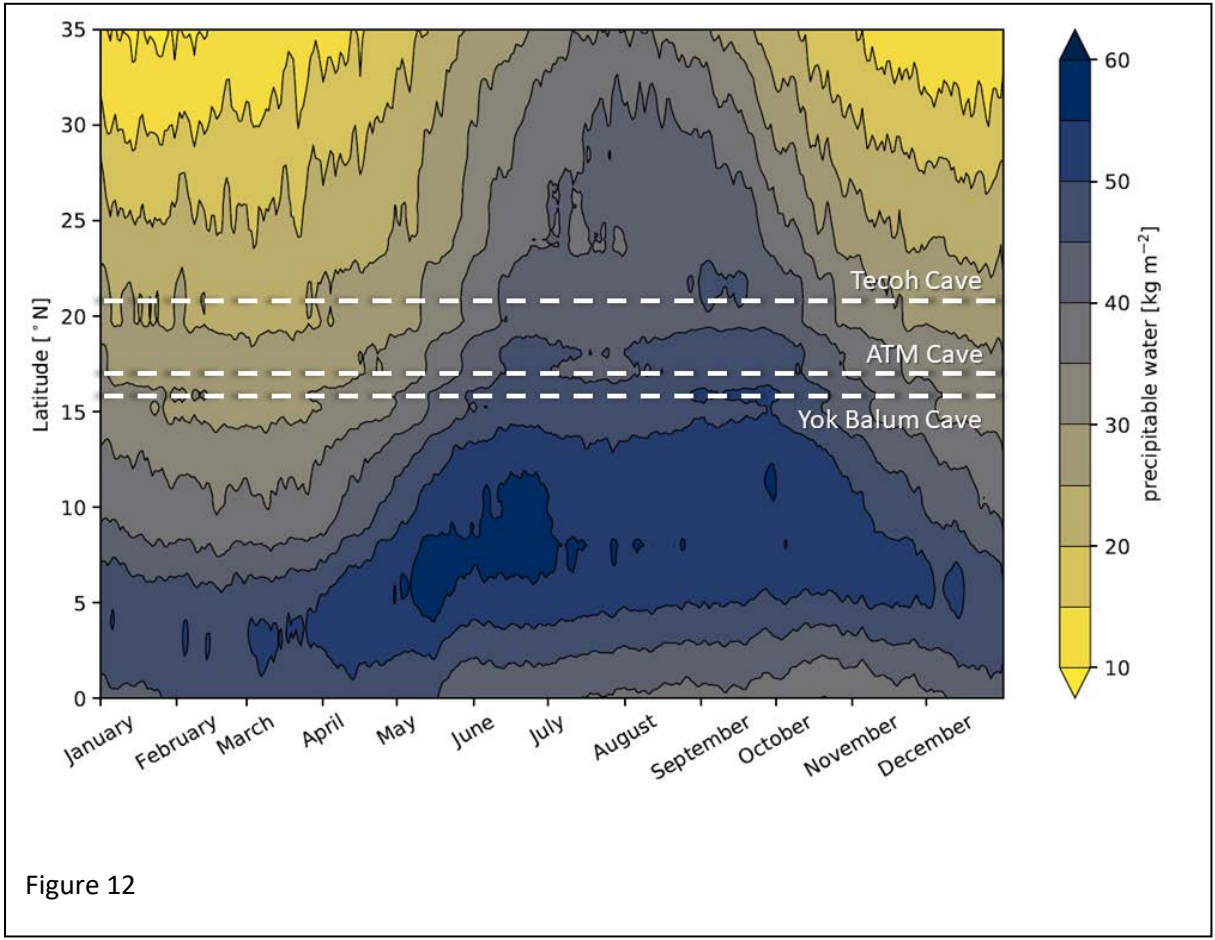

Figure 12